\documentclass[graphics,floatfix,footinbib,tightenlines,nobibnotes,aps,prb,twocolumn]{revtex4-1}
\usepackage{amsmath}
\usepackage{amssymb}
\usepackage{graphicx}
\usepackage{comment}
\usepackage{soul,xcolor}
\usepackage{bbold}
\usepackage[colorlinks=true ,urlcolor=blue,urlbordercolor={0 1 1}]{hyperref}
\usepackage{tikz}
\usepackage{braket}
\usepackage[english]{babel}
\usepackage{graphicx}
\usepackage{lipsum}
\usepackage{xr}
\renewcommand{\v}[1]{\mathbf{#1}} % \v -> vector (bf)

\newcommand{\be}{\begin{equation}}
\newcommand{\ba}{\begin{align}}
\newcommand{\ee}{\end{equation}}
\newcommand{\bea}{\begin{eqnarray}}
\newcommand{\eea}{\end{eqnarray}}
\newcommand{\beq}{\begin{equation}}
\newcommand{\eeq}{\end{equation}}
\newcommand{\beqn}{\begin{eqnarray}}
\newcommand{\eeqn}{\end{eqnarray}}

\newcommand{\bfk}{\mathbf{k}}

\newcommand{\moire}{moir\'e }

\begin{document}

\title{Stability of Anomalous Hall Crystals in multilayer rhombohedral graphene}
\author{Zhihuan Dong}
 
\author{Adarsh S. Patri}
 
\author{T. Senthil}

\affiliation{Department of Physics, Massachusetts Institute of Technology, Massachusetts 02139, USA}
 
\date{\today}

\begin{abstract}
Recent experiments showing an integer quantum anomalous Hall effect in pentalayer rhombohedral graphene have been interpreted in terms of a valley-polarized interaction-induced Chern band. The resulting many-body state can be viewed as an Anomalous Hall Crystal (AHC), with a further coupling to a weak \moire potential. We explain the origin of the Chern band and the corresponding AHC in the pentalayer system. To describe the competition between AHC and Wigner Crystal (WC) phases, we propose a simplified low-energy description that predicts the Hartree-Fock phase diagram to good accuracy. This theory can be fruitfully viewed as `superconducting ring' in momentum space, where the emergence of Chern number is analogous to the flux quantization in a Little-Parks experiment. We discuss the possible role of the \moire potential, and emphasize that even if in the moir\'e-less limit, the AHC is not favored (beyond Hartree-Fock) over a correlated Fermi liquid, the \moire potential will push the system into a `moir\'e-enabled AHC'. We also suggest that there is a range of alignment angles between R5G and hBN where a $C = 2$ insulator may be found at integer filling. 
\end{abstract}

\maketitle
\section{Introduction}
In the last few years, the Integer Quantum Anomalous Hall (IQAH) effect has been seen in a number of \moire materials\cite{chen2020tunable,serlin2020intrinsic,chen2021electrically,polshyn2020electrical,li2021quantum,foutty2023mapping}. Microscopically, typically, the electronic bands of many of these \moire materials have a well-defined Chern number that is equal and opposite in the two valleys. When the bands are nearly flat, and the total number of electrons is an odd integer per \moire unit cell, the valley (and spin, if present) degree of freedom polarizes spontaneously\cite{zhang2019nearly,bultinck2020mechanism,zhang2019twisted,repellin2020ferromagnetism}, leading to an insulating ground state with a net Chern number for occupied bands, and hence to an IQAH effect. 

The recent discovery\cite{lu2023fractional} of the IQAH (and a Fractional Quantum Anomalous Hall (FQAH)) state in pentalayer rhombohedral graphene (R5G) nearly aligned with a hexagonal Boron-Nitride substrate (R5G/hBN) does not fit this paradigm. 
Rather, the non-interacting band structure is such that, even with spin and valley polarization, the many-body state is metallic at a filling $\nu = 1$ of the moire lattice. 
However, electron-electron interaction effects, treated in a Hartree-Fock aproximation\cite{dong2023theory,zhou2023fractional,dong2023anomalous,guo2023theory}, lead to the appearance of a single-particle gap, and an insulating many-body ground state (see also Ref.~\onlinecite{kwan2023moir}). 
The occupied Hartree-Fock band has a non-zero Chern number $|C| = 1$ which provides an explanation for the observed IQAH at $\nu = 1$. 
Furthermore, numerical calculations\cite{dong2023theory,zhou2023fractional,dong2023anomalous,guo2023theory} of the many-body state at fractional filling, by considering the Coulomb interaction projected to the $\nu = 1$ Hartree-Fock band, find fractional quantum Hall states in general agreement with the FQAH found in the experiment.  The physics of the QAH in pentalayer graphene is thus quite different from the twisted Transition Metal Dichalcogenide (tTMD) system MoTe$_2$ where the first discovery of the FQAH was made\cite{cai2023signatures,zeng2023integer,park2023observation,xu2023observation}.

In this paper, we focus on the IQAH state and provide a deeper understanding of the emergence of the interaction-induced Chern band at $\nu = 1$. 
We will mostly limit ourselves to the Hartree-Fock approximation, and explain why the Chern band is stabilized. 
We will show that there is a range of alignment angles for which the non-interacting band already has Berry curvature close to $2\pi$.  
In this regime, with interactions, the Fock term dominates and both opens up a band gap and modifies the integrated Berry curvature to be exactly $2\pi$ to yield a $C = 1$ Chern insulator. We give an intuitive explanation to this phenomenon through an analogy to flux quantization in a  superconducting ring but now in momentum space.
As the alignment angle is reduced (toward the one in the devices of the experiments of Ref. \onlinecite{lu2023fractional}), the net Berry curvature of the non-interacting model within the first Brillouin zone decreases to well below $2\pi$. In this regime, the Fock term is not enough to stabilize a $C = 1$ insulator. However we show that the combination of Hartree and Fock terms suffice, and we explain the associated physics. 
Briefly, in this regime of alignment angles, the Fock term opens up a band gap and gives an insulator. 
However, the stabilization of the $C = 1$ insulator over the $C = 0$ one is due to the Hartree term. 
This is roughly because the charge distribution in the Chern insulator is more homogenous than the trivial insulator due to the impossibility of strongly localizing the electrons in a Chern band.

%\zhihuan{Do we advertise the pseudopotential here?}

 It was noted in Ref.~\onlinecite{zhou2023fractional,dong2023anomalous} (and reproduced in our own calculations) that the Hartree-Fock calculation produces a Chern insulator even in the absence of an explicit \moire potential. A natural interpretation is that the continuous translation symmetry, present in the absence of the \moire potential, is spontaneously broken to form some kind of crystalline state. However, since the Hartree-Fock description of the crystal has occupied bands with a net Chern number, this state should be viewed as an `Anomalous Hall Crystal'\cite{tevsanovic1989hall} (AHC). The \moire potential will, at the very least, pin the AHC. From this point of view, our results can be viewed as an explanation (within Hartree-Fock) for what stabilizes the AHC over either an ordinary Wigner crystal (WC) or a Fermi liquid (FL) metal. 

To capture the competition between the AHC and the WC in rhombohedral $n$-layer graphene (RnG), we propose a complementary simplified model that focuses on the symmetry indices at the MBZ corners $K_M$ and $\overline{K}_M$, in the spirit of a `Landau' theory. The ground state energy, expressed as a function of these indices, contains the key information about the AHC - WC competition.
To get the parameters for the Landau theory, we first formulate a crude treatment that focuses solely on a pseudopotential interaction between the MBZ corners. We show that although this correctly captures the Hartree term, it does not faithfully describe the effect of the Fock term in the full range of microscopic parameters. 
To remedy this, we propose a modified low-energy model that captures the physics of  the entire MBZ boundary. This model can be viewed as a superconducting ring, but in momentum space, and gives an intuitive description of the physics. This modification significantly improves our prediction of the Hartree Fock phase diagram. Similar momentum space ``superconducting" analogies have been invoked to understand Hartree-Fock physics in twisted bilayer graphene problems before\cite{Soejima2020efficient, Liu2021nematic}. 

The evolution between the Fermi liquid metal and the ordinary Wigner crystal is, of course, one of the classic problems in condensed matter physics. 
In the context of RnG, the Hall crystal is another possible phase. 
It is well known that the phase competition between the Fermi liquid and the ordinary Wigner crystal is very poorly described by Hartree-Fock theory which enormously overestimates the stability of the crystal. We might expect that a similar situation also arises in RnG with the crystalline phases (AHC or WC) much less stable than indicated by Hartree-Fock theory. 
However, we suggest that even a very small \moire potential can induce a phase transition between a correlated Fermi liquid and the Hall crystal as the latter gains commensuration energy. 
Thus, we expect the Hartree-Fock calculation to be more reliable in the presence of \moire than without. We dub the resulting IQAH state a `moir\'e-enabled Anomalous Hall Crystal'.  Obviously, in the regime of the moir\'e-enabled AHC, if we turn off the \moire potential, the system is a Fermi liquid. In this situation, the \moire potential is important in getting into the AHC state (through a first-order transition out of the Fermi liquid), but ultimately the gap in the AHC is determined by the Coulomb interaction and not by the amplitude of the periodic potential. We comment on some implications of this possibility. 

We also consider the possibility of obtaining interaction-induced Chern bands with higher Chern numbers. Indeed the naive expectation about RnG/hBN is that, for the topologically non-trivial orientation of the displacement field, the bands have (valley) Chern number $n$. Though this expectation is not borne out \textcolor{red}{\cite{chen2020tunable}}, there is a large Berry curvature in the non-interacting band that is peaked in a ring in momentum space. The size of the ring is set by the displacement field. 
Thus, with increasing twist angle, a greater portion of this Berry curvature will sit inside the \moire Brillouin zone. 
Then, if interactions open up a band gap at \moire filling $\nu = 1$, we might expect a filled band with a higher Chern number. Complicating this expectation, the non-interacting dispersion loses its flatness also at the location of the ring. 
Hence, a detailed calculation is required. We will show, within the Hartree-Fock approximation for R5G/hBN, that if the strength of the \moire potential or the Coulomb interaction strength can be tuned, then $C > 1$ insulators may be obtained at $\nu = 1$, which then raises the possibility of exploring the physics at fractional filling of a higher Chern band. 
Previously a $C = 2$ QAH state was found\cite{chen2020tunable} in R3G/hBN at $\nu = 1$ of the valence band, and a $C = 5$ insulator at neutrality in spin-proximitized  R5G at neutrality\cite{han2023large}. 

To set this work in a broader historical context, we note that a number of previous theoretical and experimental papers\cite{ghaemi2007higher,raghu2008topological,sun2009topological,polshyn2022topological,pierce2021unconventional} have explored the possibility that interaction effects can induce Chern bands and lead to an integer quantum hall effect in diverse situations. The mechanisms identified in this paper are specific to RnG, and are distinct from that in this prior literature. A particularly interesting feature of the RnG system is the possibility that, apart from time reversal, (approximate) translation symmetry is broken spontaneously as well, thereby realizing a (moire-enabled) AHC. The case of continuous translation symmetry breaking, accompanied by a quantum Hall effect, in Landau levels was discussed in Ref. \onlinecite{tevsanovic1989hall} which was itself motivated by the prior discussion of a quantum Hall effect in a Wigner crystal driven by cooperative ring exchange\cite{kivelson1986cooperative,kivelson1987cooperative}.   The coexistence of the quantum Hall effect and crystal symmetry breaking is reminiscent of the much discussed\cite{andreev1969quantum,leggett1970can} but elusive phenomenon of supersolidity in Helium-4. A recent experiment on bilayer graphene has suggested evidence for an AHC phase\cite{seiler2022quantum}. Finally, in moire systems, there are examples\cite{polshyn2022topological,pierce2021unconventional} where discrete translation symmetry of a lattice system is broken spontaneously and leads to a Chern insulator. In these examples, the folding of the Brillouin zone due to a commensurate charge density wave order captures the pre-existing Berry curvature of the bare band within the reconstructed first Brillouin zone. As we will see, the mechanism in RnG/hBN is more intricate than these examples. 

The remainder of the paper is organized as follows.
In Sec.~\ref{sec_continuum_model_begin} we describe the continuum models for R5G, and highlight the limitations of a simple and popular two-band model. Specifically, it fails to capture the realistic Berry curvature distribution at small momentum, which is critical to our understanding of the AHC.
We next overview the features of the Hartree-Fock phase diagram in Sec.~\ref{sec_hf_understanding}, for varied twist angles and displacement field energies, and provide an understanding for the interaction-driven Chern bands.
We next detail the mechanism by which the AHC is favored over the trivial WC in Sec.~\ref{sec:symm} through a careful consideration of the symmetry indices at high-symmetry points of the mini-Brillouin zone. We also provide a complementary argument in Sec.~\ref{sec_pseudopotential} via a Landau-like effective model. 
In Sec.~\ref{modifiedtheory}, we propose the aforementioned low-energy ``superconducting ring'' model which predicts the microscopic mean-field phase diagram using the non-interacting band structure.
In Sec.~\ref{sec_competing}, we discuss the competition of the AHC and the Fermi liquid state in the moir\'e-less setting within Hartree-Fock. 
In Sec.~\ref{sec_beyond}, we discuss features of the phase diagram beyond the Hartree-Fock approximation, and comment on the fate of both the \moire and non-\moire system depending on the interactions and displacement field. We emphasize the presence of a \moire-enabled AHC which expands (likely vastly) the relevance of the physics of the AHC in the phase diagram. In Sec.~\ref{sec:highC} we discuss possible routes for higher Chern bands.
Finally, we conclude in Sec.~\ref{sec_discussion} and provide some future directions of exploration.

\section{Structure of continuum band at large displacement field}
\label{sec_continuum_model_begin}

\begin{figure}
    \centering
    \includegraphics[width=0.7\linewidth]{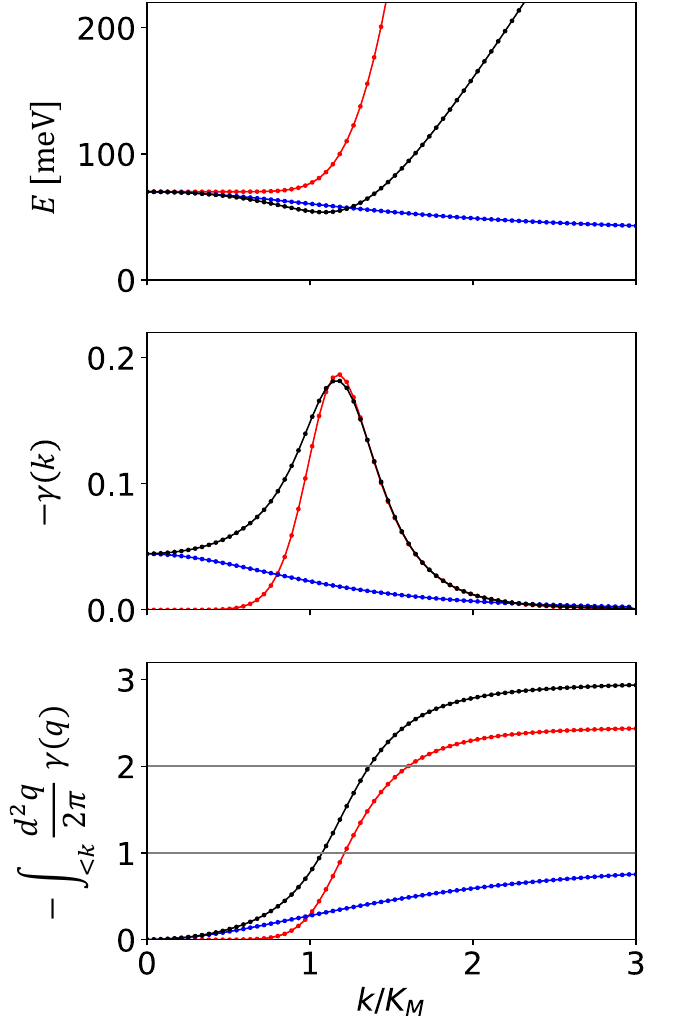}
    \caption{Dispersion (top), Berry curvature distribution (middle), and Berry flux integral (bottom) for various effective models of pentalayer graphene. 
    Black, red, and blue curves correspond to 10-band (Eq.~\ref{ten_band_model_penta} in Appendix~\ref{app_ten_band_model}), 2-band (Eq.~\ref{2band}), and 3-band models (Eq.~\ref{3band}), respectively. This plot is obtained for $u_d=-35$meV, and $\theta=0.9^\circ$.}
    \label{berry3model}
\end{figure}

Rhombohedral-stacked pentalayer graphene is the structure formed by five layers of graphene vertically aligned in the $(ABCAB)$ sequence, with the $B_l$ sublattice of layer $l$ aligned with the $A_{l+1}$ sublattice of layer $l+1$.
The continuum 10-orbital Hamiltonian (in the basis of the $A_l, B_l$ sublattices for each of the five layers; for a given valley $K/K'$ and spin $\uparrow$/$\downarrow$) involves tunneling of Dirac fermions from each graphene monolayer (with Dirac velocity $\gamma_0$) to neighboring layers via hopping processes $t_{1,2,3,4}$ (see Eq.~\ref{ten_band_model_penta} in Appendix~\ref{app_ten_band_model}) in the presence of on-site potentials and displacement field energies ($u_d$) \cite{macdonald_trilayer_2010}.
We present in Fig.~\ref{fig:ring} and Fig.~\ref{berry3model} the (conduction) continuum band structure, and the associated Berry curvature and flux of the 10-band model.

The hierarchy of energy scales associated with the inter-layer hopping processes is $t_1 > t_{3,4} \gg t_2$.
This indicates the important role of $t_1$ in the low-energy theory.
To develop an intuitive low-energy description of this continuum model, and its corresponding non-trivial topology, it is advantageous to consider various effective models that are applicable in different momenta regimes. 

We will drop the \moire potential in our discussion of the Hartree-Fock calculation, and develop an understanding of the stabilization of the AHC as a function of the electron density and the displacement field. We find it convenient to parametrize the electron density in terms of the alignment angle $\theta$ with the hBN substrate (even though we turn off the \moire potential). 
If the \moire potential were to be present, the twist angle $\theta$ determines the size of the unit cell. The charge density of interest corresponds to a lattice filling $\nu = 1$ of this unit cell. 
Consequently, $\theta$ can be used as a parametrization of the density even in the absence of \moire.

\subsection{Dangling edges of pentalayer graphene: $k^5$-two-orbital model}
A preliminary understanding of the continuum band near charge neutrality is achieved by 
focusing on the electronic occupation of the $A_1$ and $B_5$ orbitals.
The remaining eight orbitals are regarded as high-energy sites, with a low-energy model derived from perturbatively integrating out these high-energy orbitals \cite{macdonald_trilayer_2010}. 
The justification for this simplified model is that at small momentum $k$, $\gamma_0 k/t_1\ll 1$, and so the dominating interlayer hopping $t_1$ pushes all other orbitals away from zero energy by forming bonding and anti-bonding states so that the only low energy degrees of freedom are $A_1$ and $B_5$ orbitals. The effective low energy theory is obtained by performing a fifth-order perturbation with small in-plane hopping $\gamma_0k$, leading to the effective model \cite{macdonald_trilayer_2010,ho_2016, chittari_2019},
\be
    H =
    \begin{bmatrix}
    2u_d & \frac{\gamma_0^5}{t_1^4} (k_x+ik_y)^5 \\
    \frac{\gamma_0^5}{t_1^4} (k_x-ik_y)^5  & -2u_d
    \end{bmatrix}.
    \label{2band}
\ee
We note that we disregard the sub-dominant hopping momentum corrections that involve lower powers of momentum $k$.
This theory defines two momentum regimes:\\
(1) $\frac{u_d^{1/5}t_1^{4/5}}{\gamma_0} \ll k \ll \frac{t_1}{\gamma_0}$, where the off-diagonal term dominates, which leads to a $\sim k^5$ dispersion; and\\
(2) $k\ll \frac{u_d^{1/5}t_1^{4/5}}{\gamma_0}$, where the displacement field dominates, corresponds to an extremely flat band bottom (dictated by $u_d$) with a $\sim k^{10}$ dispersion.

\begin{figure}[t]
    \centering
    \includegraphics[width=\linewidth]{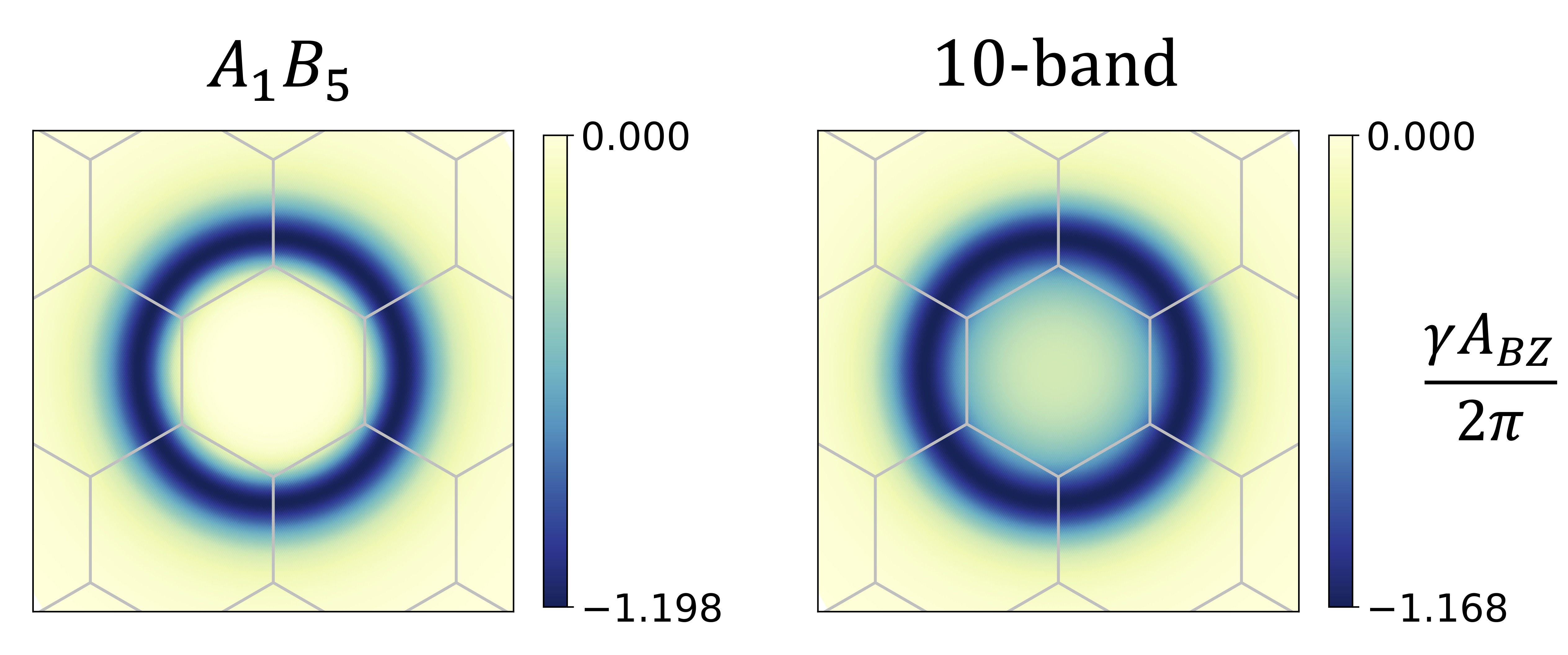}
    \caption{Berry curvature distribution for (a) the 2-orbital ($A_1B_5$) model, and (b) the 10-orbital model without warping ($t_2=t_3=t_4=0$), at $u_d=-35$meV. We mark the edge of MBZ at $\theta=0.8^\circ$ with silver lines, although no \moire potential is turned on in the calculation.}
    \label{fig:ring}
\end{figure}

The distribution of the Berry curvature associated with this two-band model is given by,
\be
%    \gamma(k)= \frac{ 25 u_d (t_1^4 / \gamma_0^5) k^8 }{(k^{10} + 4 (t_1^4/\gamma_0^5)^2u_d^2)^{3/2}}.
    \gamma(k)= -\frac{ 25 u_d (t_1^{4/5} / \gamma_0)^8 k^8 }{\Big[(t_1^{8} / \gamma_0^{10}) k^{10} + 4 u_d^2\Big]^{3/2}}.
\ee
Indeed, the corresponding total Berry flux is $-5\pi$ within the first Brillouin zone (though there is no gap), which is distributed in a ring-like region about $k=0$ (see Fig.~\ref{fig:ring}(a)). 
The radius of this ring is given by the transition momentum scale $k_{ring}\sim\frac{u_d^{1/5}t_1^{4/5}}{\gamma_0}$, where the band starts to become highly dispersive.

In the presence of the \moire potential, there is another momentum scale, which depends on the twist angle $\theta$. 
We will characterize it with $K_M$ the momentum at the corner of the \moire Brillouin zone (MBZ). 
At the range where the IQAH is observed in current experiments($\theta\sim 0.77^\circ$, $u_d\sim -35$meV) $k_{ring}$ is slightly above $K_M$ (as seen in Fig. \ref{fig:ring}(a)).
This indicates that the majority of the Berry curvature arising from the $A_1-B_5$ hybridization lies outside the first MBZ.

%\subsection{Effective model for flat conduction band bottom}
\subsection{Failure of $k^5$-two-orbital model to capture flat band bottom}

The limitation of the $k^5$-two-orbital model is brought to the fore as the displacement field energy increases.
We recall from the above discussion that at relatively weak displacement field energies ($|u_d|<10$meV) and momenta, the band bottom is relatively flat.
As such, the $k^5$-two-orbital model (in regime (2)) faithfully captures the features of the 10-orbital model.
As the displacement field increases (see Fig. \ref{berry3model}) and the flat-band region expands to envelop the size of the \moire BZ, the band-bottom (i.e. near $k=0$) acquires an appreciable dispersion. 
This appreciable dispersion is missed by the above simplistic two-orbital model.
To see this, one notices that 
the hybridization between $B_5$ and $B_4$ is linear in $k$ (see the 10-orbital model in Eq.~\ref{ten_band_model_penta} in Appendix~\ref{app_ten_band_model}) and can become much stronger than the effect of the $k^5$ term at small $k$. 
Therefore, it is more appropriate to begin with the Hamiltonian projected to include this nearest orbital i.e. in the $B_5$, $A_5$, and $B_4$ orbitals\footnote{In the following discussion, we make the simplification of ignoring other on-site potentials present in the real system; they are easily incorporated when more precision is needed.}

%In Fig. \ref{berry3model}, we demonstrate the continuum band structure calculated with 10 orbitals (the Hamiltonian of the full-10 band model is presented in Appendix \ref{app_ten_band_model}). 
%At a relatively weak displacement field ($|u_d|<10$meV) the band bottom is flat as expected from the discussion above. 
%As the displacement field increases (see Fig. \ref{berry3model}) and the flat-band region expands to envelop the size of the \moire BZ
%about the size of the \moire BZ, the band-bottom acquires dispersion. %again. 
%In fact, at strong displacement fields, the $k^5$ theory fails to be a good approximation, especially for the regime (2). 
%To see this, we note that the hybridization between $B_5$ and $B_4$ is linear in $k$ and can become much stronger than the effect of $k^5$ term at small $k$. Therefore we write down the Hamiltonian projected to $B_5$, $A_5$, and $B_4$ orbitals

%%%
\be
    H_3 =
    \begin{bmatrix}
    2u_d & \gamma_0 (k_x+ik_y) & 0 \\
    \gamma_0 (k_x-ik_y)  & 2u_d & t_1\\
    0 & t_1 & u_d
    \end{bmatrix}.
    \label{3band}
\ee
%%%
Diagonalizing this Hamiltonian in the limit of small $k$ and dominating $t_1$, we find the eigenstates: 
%If we diagonalize the dominating $t_1$ term first and find the eigenstates are 
$\ket{B_5}, \ket{A_5B_4^+}=\frac{1}{\sqrt{2}}(\ket{A_5}+\ket{B_4})$, and $\ket{A_5B_4^-}=\frac{1}{\sqrt{2}}(\ket{A_5}-\ket{B_4})$. 
Written in this eigenbasis basis, the Hamiltonian becomes
\be
    \tilde{H}_3 =
    \begin{bmatrix}
    2u_d & \frac{\gamma_0 (k_x+ik_y)}{\sqrt{2}} & \frac{\gamma_0 (k_x+ik_y)}{\sqrt{2}} \\
    \frac{\gamma_0 (k_x-ik_y)}{\sqrt{2}}  & t_1+\frac{3u_d}{2} & \frac{u_d}{2}\\
    \frac{\gamma_0 (k_x-ik_y)}{\sqrt{2}}  & \frac{u_d}{2} &-t_1+\frac{3u_d}{2}
    \end{bmatrix}
\ee
At $k=0$, the lowest conduction band state is $\ket{\psi^0}=(1,0,0)=\ket{B_5}$, as expected. 
For small $k$, the $k$-dependent matrix elements can be treated within perturbation theory.
The leading correction to the eigenstate of the lowest conduction band is,
%Now consider perturbing in terms of $k$, the leading correction is
\begin{align}
\label{psi_perturb}
    \ket{\delta\psi_1} &= -\frac{\gamma_0 k^*}{2} \left( \frac{u_d}{t_1^2-u_d^2/4}\ket{A_5} +\frac{2t_1}{t_1^2-u_d^2/4}\ket{B_4}\right) \nonumber\\
    &\sim -\frac{\gamma_0 k^*}{t_1}\ket{B_4}
\end{align}
and its corresponding energy correction is,
\be
\label{E_perturb}
    \delta E_2 = -\frac{\gamma_0^2|k|^2}{2} \left( \frac{1}{t_1-u_d/2} -\frac{1}{t_1+u_d/2}\right) \sim -\frac{u_d\gamma_0^2 |k|^2}{2t_1^2}.
\ee
The leading correction to the dispersion in Eq.~\ref{E_perturb} suggests a negative effective mass that scales with displacement field.
As such, the band bottom is not as flat in the large displacement field energy as the $k^5$-two-orbital model would seem to indicate.
We present in Fig. \ref{berry3model} the conduction band bottom dispersion for the $k^5$-two-orbital model and the above perturbation theory modified effective three-orbital model  (with the 10-orbital model's dispersion shown for comparison).
As seen, the $k^5$-two-orbital model is `too flat' in the small momentum region; the effective three-orbital has the required dispersion, consistent with the full 10-orbital model.

%which is consistent with Fig.~\ref{berry3model}.

In addition to the modified dispersion, the correction to the $\ket{B_5}$ eigenstate by Eq. \ref{psi_perturb} indicates that the Berry curvature in the small $k$-regime resembles that of a massive Dirac cone. 
This entails a Berry phase of approximately $-\pi$ distributed in the small region of $k\sim t_1/\gamma_0$.
This is corroborated by the full 10-orbital model, where $\gamma_{BZ}$, the Berry flux inside the MBZ, is slightly stronger than $-\pi$ (see Fig.~\ref{berry3model}); recall that at the larger displacement fields discussed here, the `ring' of Berry curvature lies just outside the MBZ.
While the ring-shaped feature at larger momentum arising from $B_5-A_1$ mixing also contributes to $\gamma_{BZ}$, the described $-\pi$ flux is predominantly due to the massive Dirac cone physics at small $k$ due to $B_5-B_4$ mixing. 
In the following section, we will see that this massive Dirac cone feature is crucial since the $\gamma_{BZ}$ of the non-interacting band, to a large extent, determines the Hartree-Fock phase diagram.

%Besides, Eq.~\ref{psi_perturb} also suggests that the quantum geometry in the small $k$ regime resembles that of a massive Dirac cone, namely there should be roughly $\pi$ Berry phase smoothly distributed within the region of $k\sim t_1/\gamma_0$. 
%In the free continuum band derived from the full 10-orbitals (see Fig.~\ref{berry3model}), we find the Berry flux $\gamma_{BZ}$ inside the 1st BZ to be slightly above $\pi$. This combines the contribution of the massive Dirac cone physics at small $k$ due to $B_5-B_4$ mixing, as well as the ring-shaped feature at larger momentum arising from $B_5-A_1$ mixing.

%\section{Phase diagram}
\section{Overview of the Hartree-Fock Phase Diagram}
\label{sec_hf_understanding}

\begin{figure}
    \centering
    \includegraphics[width=\linewidth]{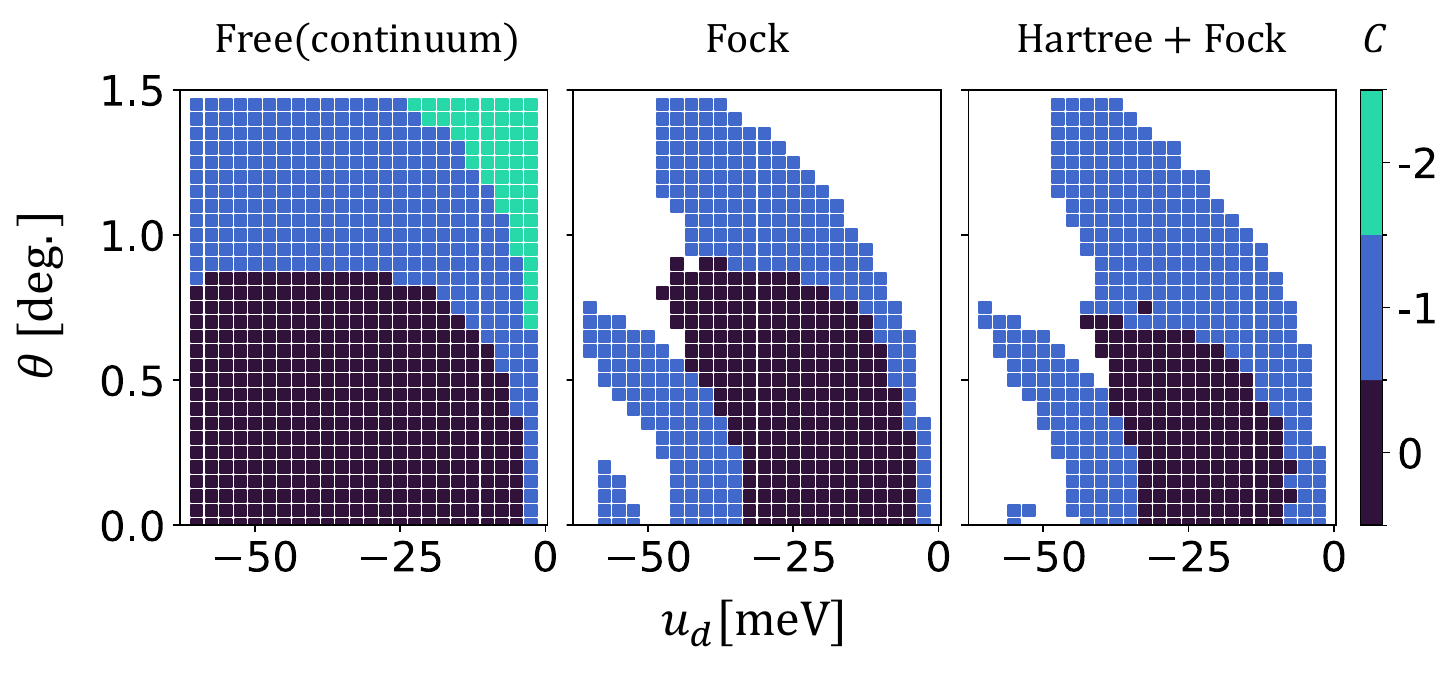}
    \caption{Phase diagrams in $\theta-u_d$ space. In (a) The colors show the closest integer to $\gamma_{BZ}/2\pi$, where $\gamma_{BZ}$ is defined in the main text. Colors in (b) and (c) mark the Chern number of mean field bands, obtained with Fock term only and full Hatree-Fock, respectively. Here the metallic region is left uncolored.
    Comparing (a) and (b), the boundary between $C=0$ and $C=-1$ region in the Fock phase diagram closely tracks that of the non-interacting phase diagram. The comparison between (b) and (c) shows that the Hartree term extends the stability of the  $C=-1$ Hall crystal. 
    These results are obtained for the 10-orbital model with no warping term, and with the \moire potential turned off. The angle $\theta$ should then be viewed as a parametrization of the charge density. This result is obtained with $24\times24$ momentum mesh and projecting to 4 lowest conduction bands. In Appendix~\ref{app:7band}, we show the results including 7-bands. These results are consistent except for the regime of large $u_d$ and small $\theta$, where the ``Mexican hat'' dispersion is deep enough so that the wavefunction at $\Gamma_M$ point is a linear combination of 6 MBZ neighboring the first MBZ, thereby it is crucial to include 7 bands.}
    \label{hf_pd_compare}
\end{figure}

In this section, we overview the phenomenology of the mean-field findings of pentalayer graphene and explain the physics for some parts of the phase diagram. 
In particular, we focus on the conduction band in the regime of strong displacement field ($u_d\sim -35$meV) relevant to current experiments on R5G. We consider the evolution as twist angle $\theta$ ({\em i.e.} the charge density) decreases, and ignore the \moire potential. 
The Hartree-Fock and Fock phase diagrams are shown in Fig.~\ref{hf_pd_compare} \footnote{The methodology of the Hartree-Fock calculations follow those in Ref. \onlinecite{dong2023theory}}. 
We focus on a simplified model with no warping $t_2=t_3=t_4=0$; the warping effects are readily included and do not affect the physical understanding we develop. 

\vspace{0.3cm} 
\underline{{\em Extremal twist angle $\theta>1.5^\circ$:} }
In this limit, $K_M>k_{ring}$, and so the majority of the continuum band Berry curvature is contained within the first MBZ. 
We recall that $k_{ring}$ also marks the momentum scale at which the continuum band becomes dispersive. 
As a result, given that the band gap opening at the mean-field level is bounded (from above) by the scale of interaction $U\sim 20$meV, the energy of the second conduction band at the $M_M$ point may drop well below the energy of the first conduction band at $K_M$, which leads to a metallic phase even when the translation symmetry is spontaneously broken.
%\adarsh{why M point in particular?} \zhihuan{because $M$ is the smallest momentum on BZ boundary. So the second conduction band is likely to be lowest at $M$}

\vspace{0.3cm} 
\underline{\em Large twist angle $0.9^\circ<\theta<1.5^\circ$:} In this case, where the twist angle is larger than the experimental regime, the MBZ boundary intersects the Berry curvature ring, where the $-5\pi$ Berry flux is concentrated.
We find the Berry flux enclosed by the first MBZ varies between $-3\pi$ and $-\pi$ as $\theta$ decreases. 
With Fock term alone turned on, the band gap opens, and we find the lowest conduction band has $C=-1$, which is exactly the Chern number expected from rounding $\gamma_{BZ}/2\pi$ to its closest integer. 
The inclusion of the Hartree term does not modify the result qualitatively.

We note that in this regime an AHC with a higher Chern number is potentially possible. For this to happen, we need the twist angle to be large enough to enclose more than $-3\pi$ flux, but at the same time not so large that the band becomes too dispersive to be gapped out by interaction. Whether or not such a higher Chern AHC exists then depends on details of the model. 
Within the current simplified model, no $|C|>1$ phase is observed in Fig.~\ref{hf_pd_compare}(c).
Nonetheless, it is conceivable that some modification to the non-interacting model parameters (such as including trigonal warping, or tuning the strength of the Coulomb interaction) can favor the higher Chern AHC. In addition, we can also contemplate tuning the strength of the \moire potential to stabilize an interaction-induced higher Chern band.   We will explore these phenomena in Sec.~\ref{sec:highC}.

\vspace{0.3cm} 
\underline{\em Intermediate twist angle $0.6^\circ<\theta<0.9^\circ$:} This is the regime most relevant to existing experiments\cite{lu2023fractional}. 
Due to the smaller MBZ, the enclosed Berry flux in the first BZ is less than $-\pi$.
%Now with a smaller MBZ, the Berry flux $\gamma_{BZ}$ now drops below $\pi$. 
With the Fock term alone, we find a trivial insulator (WC), consistent with the smaller Berry flux in the non-interacting band structure. However, with the inclusion of the Hartree term, the $C=-1$ AHC state is once again favored (as seen in Fig.~\ref{hf_pd_compare}). We emphasize that this Hall crystal state is not a straightforward expectation from the non-interacting theory. 
In Sec.~\ref{sec:symm}, we will describe a mechanism, according to which the Hartree term generically (for various displacement fields, number of graphene layers, etc.) favors $|C|=1$ AHC at a moderate interaction strength (strong enough to gap out the band, but not as strong compared to the dispersion in higher MBZs, which is the regime most relevant to the experiments).

\vspace{0.3cm} 
\underline{\em Small twist angle $\theta<0.6^\circ$:} 
Finally, at a very small twist angle, 
the MBZ is shrunk significantly and is unable to enclose any substantial Berry curvature.
%the first MBZ shrinks so much that most of the Berry curvature is excluded. 
As a result, the low energy degrees of freedom in the continuum band are almost entirely trivial and featureless i.e. the Bloch function is almost completely polarized to $\ket{B_5}$. 
With the spontaneous breaking of translational symmetry, a Wigner crystal develops.
%Then the interaction leads to a Wigner crystal just like in the jellium model.

\vspace{0.3cm} 
\underline{\em Small $\theta$ and higher displacement field $|u_d|>40$meV:} At a small twist angle, the ring-like feature is in the second MBZ. Naively one may conclude that the Chern number will be zero since $|\gamma_{BZ}|<\pi$. However, the situation is more subtle as seen in Fig.~\ref{hf_pd_compare}(c). According to Eq.~\ref{E_perturb}, a large displacement field leads to strong negative dispersion at small $k$. As a result, the lowest conduction band is now mainly formed by the second MBZ, which contains a significant part of the Berry curvature ``ring". Then the Berry flux in the second MBZ may well be close to $-2\pi$, which will get quantized to $-2\pi$ when the Fock term is turned on. So a $C=-1$ band is again expected at the level of the mean-field calculation.

\section{Role of Hartree term: a physical picture}
%\section{symmetry index and density profile of Bloch function}
\label{sec:symm}

The prevalence of the $|C|=1$ AHC in the phase diagram of Fig.~\ref{hf_pd_compare} warrants an explanation for the mechanism by which the Coulomb interaction favors $|C|=1$ AHC over $C=0$ WC. 

As noted above, when the underlying band has a total Berry curvature close to $2\pi$ within the area of the first Brillouin zone, the Fock term alone yields a Chern-1 band, {\em i.e}, the Berry curvature is rounded to exactly $2\pi$ in the Hartree-Fock band. We will explain this phenomenon in a later section. 
In this section, however, we focus on the role of the Hartree term in selecting the AHC. 
To examine it in isolation, we compare the mean-field phase diagrams with both Hartree and Fock and Fock-only in Fig.~\ref{hf_pd_compare} (b) and (c).
The inclusion of the Hartree term expands the $|C|=1$ region of the phase diagram to include the experimentally relevant twist angle $\theta\approx0.77^\circ$ and displacement field $u_d\approx 30$meV.
%As discussed in the previous section, 
It is therefore essential to understand the mechanism by which the Hartree term favors the AHC over the WC. %intermediate twist angle regime, where the Hartree term is needed to stabilize the AHC over WC. 
%This is also the regime where currently experimental observations of IQAH are made.

We recall from topological band theory \cite{hughes2011inversion,fang2012bulk} that the Chern number is determined by the symmetry indices at the invariant momentum under the point group. 
In R5G, the corresponding point group is $C_3$, with the invariant momentum $\Gamma_M$, $K_M$ and $\bar{K}_M$. 
Therefore the real question is how the interaction selects a particular set of symmetry indices $(I_{\Gamma_M}, I_{K_M}, I_{\bar{K}_M})$. 

%To further simplify the problem, 
We first note that the Bloch function at $\Gamma_M$ almost completely comes from the first MBZ, and is thus almost fully polarized to the $B_5$ orbital -- the rapidly increasing kinetic energy beyond the first MBZ prohibits substantial hybridization with higher MBZs. 
Therefore, $I_{\Gamma_M}=0$ if we define the $C_3$ axis through the $B_5$ site, which will be our convention. This allows us to focus on the remaining symmetry indices $I_{K_M}$ and $I_{\bar{K}_M}$, and ask how they affect the Hartree energy. At these two momenta, the kinetic energy again makes higher MBZs irrelevant and only allows significant hybridization among three equivalent corners of the first MBZ.

%To be specific, the eigenstate at $K_M$ is a superposition of the state at the three equivalent $K$ points; the eigenstate can be simply denoted by the three-component vector $\{\alpha_1, \alpha_2, \alpha_3\}$.
%The $C_3$ operation leads to the cyclical permutation of the coefficients ${\alpha}_i \rightarrow {\alpha}_{i+1}$, with a corresponding eigenvalue of this permutation being $-1, 0, 1$.
%Thus, we can compactly represent a Bloch state with $C_3$ symmetry index $L$ as having the coefficients $\alpha = \{1, \omega^L, \omega^{2L}\}$.

To develop some intuition for the Hartree energy, it is instructive to consider the wavefunction amplitude in real space.
%We find there is a connection between the density profile of the Bloch function and the symmetry index at these two high symmetry momentum.
The full Bloch function in Eq.~\ref{eq_bloch_fun_main} possesses two levels of structure
\begin{equation}
    \psi_{\v K_M}(\v r, s) = \sum_{\v G} \alpha_{\v G} e^{i\v G \cdot \v r} u_{s}(\v K_M+\v G).
    \label{eq_bloch_fun_main}
\end{equation}
First is the structure in 10 atomic orbitals $u_s$, which controls the profile at the scale of the pentalayer graphene unit cell. 
The second is from the hybridization of different MBZ $\alpha_{\v G}$, controlling the profile at the scale of the moir\'e unit cell. 
For simplicity, let us consider the case where the 10-orbital Bloch functions of $K_M$ and $\bar{K}_M$ get polarized to $B_5$. Then the angular momentum (or $C_3$ index) is completely attributed to the hybridization of three plane wave components at the three equivalent $K_M$ or $\bar{K}_M$ points.

\begin{figure}[t]
    \centering
    \includegraphics[width=1\linewidth]{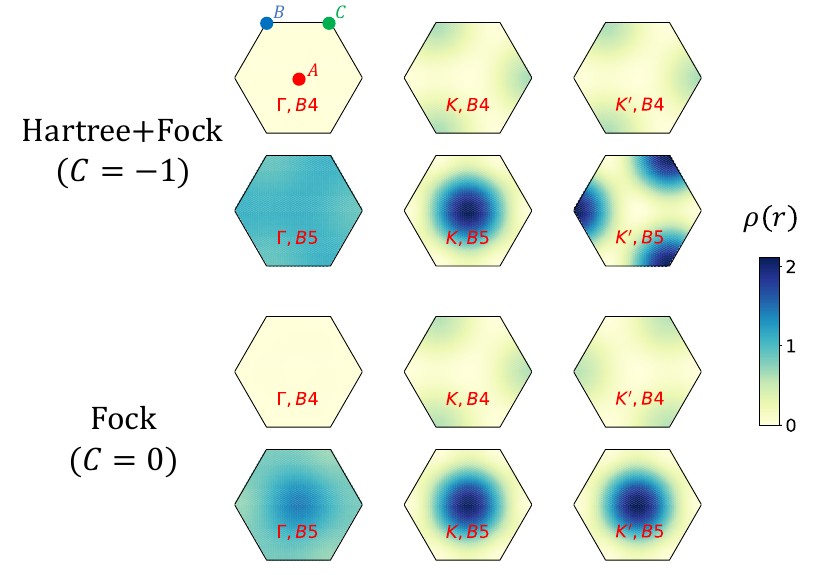}
    \caption{Real-space $B_4$ and $B_5$ density profile for Bloch functions at high symmetry momentum. 
    With the Hartree term included, the $B_5$ orbitals repel and the density profile is spread out to different regions of the moir\'e unit cell, while the $B_4$ orbitals overlap. 
    Without the Hartree term, the Fock energy is reduced by overlapping $B_5$ orbitals. 
    These results are obtained for $\theta=0.7^\circ$, $u_d=-30$meV, and with a very weak moire potential $C_{AA}=C_{BB}=1$meV to pin the wavefunction to high symmetry positions in the unit cell.}
    \label{denisty_profile_main}
\end{figure}

Now, we define the real-space hexagonal unit cell centered at the $C_3$ axis. The Wyckoff positions are $A,B,C$, respectively corresponding to the center and two corners of the unit cell (see Fig.~\ref{denisty_profile_main}). Since the wavefunction is symmetric under $C_3$, the peak of the Bloch function must be on $A$, $B$, or $C$. For a wavefunction with an angular momentum of $0$ under $C_3$, its density may peak at $A$, the symmetry center. (In fact, this is the only possibility since the Bloch function at $K_M$ and $\bar{K}_M$ must vanish at two of the three Wyckoff positions. See Appendix~\ref{app:density_profile}.) In contrast, for a wavefunction with nonzero angular momentum under the $C_3$, the wavefunction must be zero at $A$. The intuition is that the angular momentum pushes density away from the symmetry center, just like the effect of centrifugal force. In Appendix~\ref{app:density_profile}, we show that there is a one-to-one correspondence between the peak position $R_s =0,1,2$ corresponding to $A,B,C$, and the $C_3$ index $I_s = 0,1,2$ at a MBZ corner $K_M$ and $\bar{K}_M$ labeled by $s = +1$ and $-1$,
\begin{equation}
    R_s \equiv sI_s \text{ (mod 3)}.
\end{equation}
We also discuss the general scenario without $B_5$-polarization in Appendix \ref{app:density_profile}. Thus, the Chern number is given by,
\begin{equation}
    C \equiv I_{K_M} + I_{\bar{K}_M} \equiv R_{K_M}-R_{\bar{K}_M} \text{ (mod 3)}.
\end{equation}
Since the Hartree term always tends to keep the electrons apart, $R_{K_M}\neq R_{\bar{K}_M}$ or $C\neq 0$ state always gets favored by Hartree energy.
In Fig.~\ref{denisty_profile_main} we demonstrate this picture by plotting the density profile of Bloch function for a $C=-1$ AHC and a $C=0$ WC.

Finally, it is important to note that our above arguments are valid in a relatively narrow range of twist angle $0.6^\circ<\theta<0.9^\circ$.
%Away from this window, our approximations stop being valid. 
When $\theta$ is too large, the first MBZ is no longer completely within the flat region. Therefore at MBZ boundaries, the Bloch function no longer has its weight concentrated on $B_5$ orbitals. 
On the other hand, at a small $\theta$ and relatively large $u_d$, the Bloch functions near $\Gamma$ may acquire strongly non-uniform density profiles in real space, since $c_{\v k}$ and $c_{\v k+\v G}$ are not far away in energy and can hybridize strongly. 
In this case, it is no longer justified to consider the subsystem of only $K$ and $\bar{K}$ points.

\section{Simplified ``Landau" model for competing insulators}
\label{sec_pseudopotential}

In this part, we reformulate the story in the Sec.~\ref{sec:symm} in a more elegant way, in the spirit of a Landau theory. We will focus on the interplay between Bloch states at MBZ corners $K_M$ and $\bar{K}_M$, and show explicitly how the $C_3$ eigenvalues at these two high-symmetry points are chosen to be non-trivial by the Coulomb interaction.

For any insulating state, the filling is $n_k=1$ at both $K_M$ and $\bar{K}_M$ points. Therefore, at the mean-field level and assuming a $C_3$-preserving ansatz, the low-energy degrees of freedom describing the competition between AHC and WC are two angular variables (classical rotors), $\theta_1$ and $\theta_2$, associated with the two MBZ corners $K_M$ and $\bar{K}_M$. Each rotor variable may take a value from $\{0, 2\pi/3, 4\pi/3\}$, corresponding to $C_3$ symmetry index of 0, 1, and 2. Note that, in our setting, the $\Gamma_M$ point of the MBZ has a trivial $C_3$ index. Thus it is sufficient to focus on the two MBZ corners $K_M$ and $\bar{K}_M$. 

Here we would like to add two remarks: (1) Although $C_3$ indices are only well-defined at the high-symmetry points $K_M$ and $\bar{K}_M$, their values also determine the surrounding measure-nonzero regions since the Bloch function is continuous in $k$-space. Therefore, it is conceivable that these indices may be associated with the energy of many-body state. (2) Although the $C_3$ indices cannot directly be thought of as order parameters since they are not associated with symmetry breaking, they carry sufficient information to distinguish the WC and AHC phases. Therefore, as we focus on the competition between the two phases, we can still follow the spirit of Landau, by treating $C_3$ indices $\theta_{1,2}$ as the relevant ``macroscopic" variables for our simplified model, and then study how the symmetry of the system constrains the ground state energy as a function of $\theta_{1,2}$. 

Note that the $C_3$ index is defined modulo 3. The angular  nature of $\theta_{1,2}$ demands that the ground state energy be invariant under $\theta_1\rightarrow\theta_1+2n\pi$, $\theta_2\rightarrow\theta_2+2m\pi$, which constrains the effective model to be of the form of a 2-site and 3-state classical Potts model,
\begin{align}
    H_{\text{eff}} =& - J_+ \cos(\theta_1 + \theta_2 -\phi_+) - J_- \cos(\theta_1 - \theta_2 - \phi_-) \nonumber \\
    & + h_1 \cos(\theta_1-\phi_1) + h_2 \cos(\theta_2-\phi_2).
\end{align}
In the absence of \moire potential, the Hamiltonian is further constrained by a translational symmetry that translates the center of the hexagonal \moire unit cell to its corner (see Fig.~\ref{fig_translation}). This translation alters the $C_3$ indices at two MBZ corners in opposite directions, namely,
$\theta_1\rightarrow \theta_1+2\pi/3$, $\theta_2\rightarrow \theta_2-2\pi/3$. The proof is detailed in Appendix~\ref{symmetry}. Therefore, the only term allowed by symmetry is
\be
\label{GL2}
    H_{\text{eff}} = -J \cos(\theta_1 + \theta_2 - \phi) = -J \cos\left(\frac{2\pi C}{3} - \phi\right)
\ee
We note that in more generic settings, where all high symmetry momenta are ``activated" (meaning hybridization between $\v k$ and $\v k+\v G$ is strong), the model should involve three sites, which may have richer physics due to frustration. But in this work, we will stick to this simple and experimentally relevant two-site model. We emphasize again that the simplified model Eq.~\ref{GL2} is different from the standard Landau framework.  This is not a theory for low energy fluctuations;  rather it only captures the competition between the 9 local minima of energy topography, which appear when $C_3$ symmetry is enforced. In the usual Landau framework, high-energy fluctuations are integrated out to generate a low-energy effective Landau energy function. Here, in the same spirit,  we find energy minimum in the space of the  $C_3$ indices.

\section{Approximate calculations of ``Landau theory" parameters using simplified models}
\label{modifiedtheory}
In this section, we discuss methods to extract the Landau theory parameters from microscopics. We begin with the simplest treatment, which only considers the MBZ corners, and study the pseudopotential that directly couples the respective ``order parameters'' $\theta_{1,2}$. We show this treatment is insufficient to understand the microscopic phase diagram, as it fails to account for the role of the MBZ edges. In fact, the interaction mediated by the connecting edges dominates over the direct coupling between corners. We demonstrate this through a modified low-energy model, which resembles a superconducting ring in momentum space.

\subsection{Pseudopotential interactions for MBZ corners}
\label{sec:pseudo}
\begin{figure}[h]
    \centering
    \includegraphics[width=\linewidth]{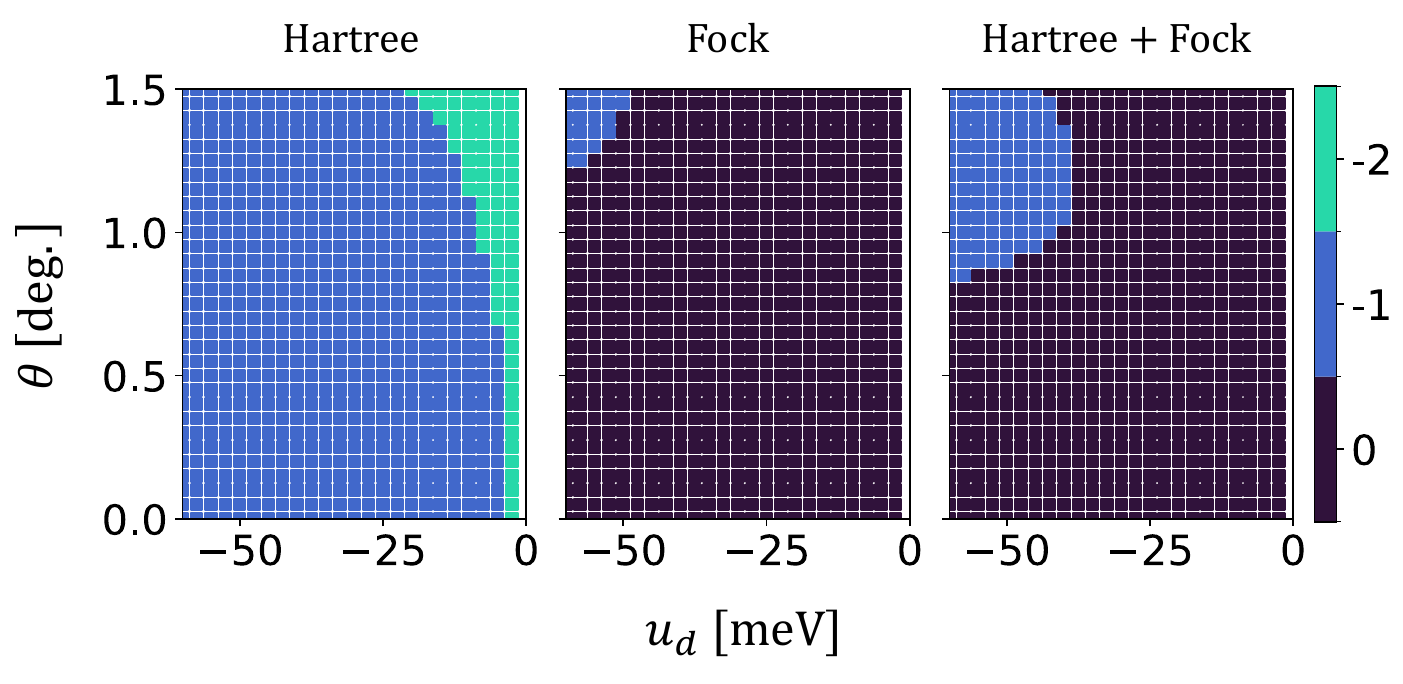}
    \caption{Hartree, Fock and Hartree+Fock phase diagrams based on the simplified pseudopotential Eq.~\ref{pseudo_final_main}.}
   \label{pd_corner_main}
\end{figure}

Focusing on the direct interaction between MBZ corners, we derive the Landau parameters $J$ and $\phi$ within Hartree-Fock theory.
This procedure amounts to rewriting the interaction in terms of pseudopotential in angular momentum channels for the MBZ corners.
The resulting Hartree-Fock energy is
\begin{align}
    H_{HF} = -\frac{2}{3} \Re\left[\left(-V(G)\lambda_{K_0,K_1}^2+ V(K)\lambda_{K_0,\bar{K}_1}^2\right) e^{i(\theta_1+\theta_2)}\right]
    \label{pseudo_final_main}
\end{align}
where $V(\v q)$ is the interaction amplitude, and $\lambda_{\v k,\v k'} = \langle u_{\v k} | u_{\v k'} \rangle$ is the form factor for the non-interacting continuum band. $\vec G$ is a vector that connects two $C_3$ related $K$-points in the MBZ while $\vec K$ is a vector that connects two adjacent corners of the MBZ. These two terms come from Hartree and Fock, respectively. Eq.~\ref{pseudo_final_main} leads to a Chern number
\begin{align}
    C \equiv -\frac{3 \arg\left(-V(G)\lambda_{K_0,K_1}^2+ V(K)\lambda_{K_0,\bar{K}_1}^2\right)}{2\pi} \text{ (mod 3)}
    \label{pseudo_Chern}
\end{align}
We leave the details of this calculation to Appendix~\ref{app_alternative_gl_model}.
Using this, we confirm the intuition from Sec.~\ref{sec:symm} that the Hartree term favors AHC over WC throughout the phase diagram (see Fig.~\ref{pd_corner_main}(a)), which explains the shift of phase boundary between Fig.~\ref{hf_pd_compare}(b) and (c).

However, this model turns out to make poor predictions for the role of Fock terms. As we show in Fig.~\ref{pd_corner_main}, the predicted Fock and Hartree-Fock phase diagram from this simplified model is far from the microscopic mean-field calculations in Fig.~\ref{hf_pd_compare}(b) and (c).

\subsection{Modified treatment for the Fock term: superconducting ring in momentum space and Berry curvature rounding}
%The previous section sets a goal: we aim for a low-energy theory that predicts the mean-field phase diagram from the non-interacting band structure. 

%The pseudopotential model in the Sec.~\ref{pseudo} fails to (1) describe the correct phase diagram (2) account for 
The pseudopotential model in Sec.~\ref{sec:pseudo} fails to correctly describe the microscopic phase diagram. 
Meanwhile, we have not yet provided a concrete explanation for the Berry curvature rounding, which we invoked to explain the Fock phase diagram (Fig.~\ref{hf_pd_compare}(b)). In this section, we kill two birds with one stone by proposing a modified low-energy model, which enables us to derive the phenomenon of Berry curvature rounding. Therefore this modified model captures the relevant physics and predicts the microscopic Hartree-Fock phase diagram to a significantly improved precision.

To make progress, we reflect on the approximations in Sec.~\ref{sec:pseudo} and Appendix~\ref{app_alternative_gl_model}. This treatment is crude in the sense that we focused on the MBZ corners, and approximated the Bloch function in their neighborhood to be uniform. This is incorrect since the microscopic low energy degree of freedom is the phase of the crystalline order parameter $\Psi_{\v G}(\v k)=\langle c^\dagger_{\v k} c_{\v k+\v G}\rangle$ which is not restricted to the corner regions, but instead remains appreciable around the entire MBZ boundary.
In the approximation of Sec. \ref{sec:pseudo}, the small momentum scattering contributes completely trivial Fock energy, i.e. does not lift the degeneracy between AHC and WC. This statement is again risky: the Fock term is in fact dominated by the small-$q$ scattering, since both the Coulomb potential and the form factor may decay rapidly with $q$.
\begin{figure}[h]
    \centering
    \includegraphics[width=\linewidth]{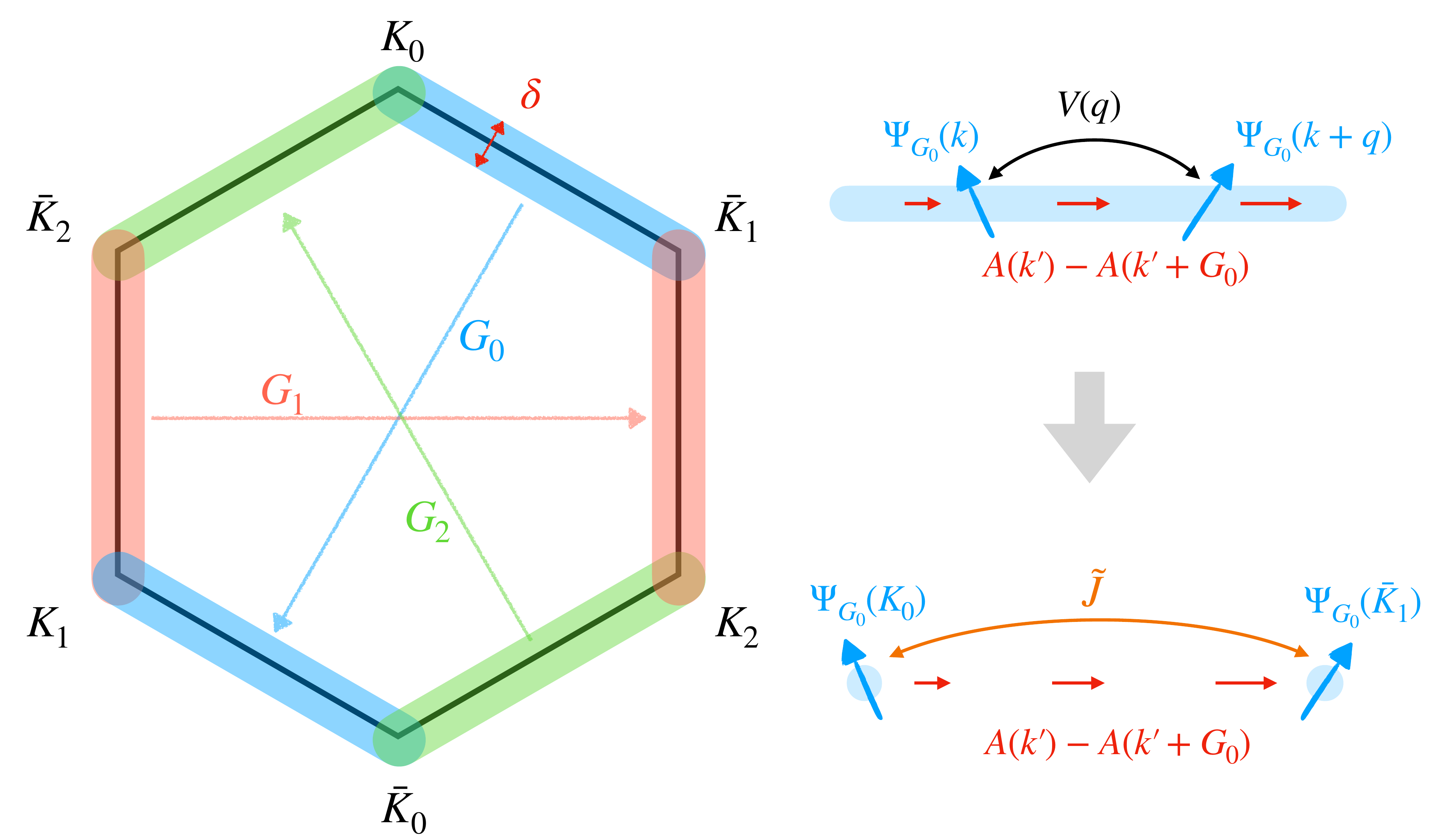}
    \caption{Superconducting ring model. (a) The crystalline order parameter is distributed in a ring-shaped region in momentum space. The low energy theory is effectively a superconducting ring subject to a background magnetic field, defined by the Berry curvature of the unfolded band. (b) The interaction between corners is generated by integrating out the ``superconducting wire'' connecting them.}
   \label{fig:SCring}
\end{figure}

In the following, we propose a modified effective model to study the competition between AHC and WC by accounting for all the edges of MBZ.  This model demonstrates the crucial role of small-$q$ scattering. We show that the model can be viewed as a momentum space analog of a superconducting ring in a magnetic field which gives an intuitive understanding of the Berry curvature rounding phenomenon.

Within each edge, there is one specific order parameter $\Psi_{\v G}(\v k)$, which is treated as an XY field on a 1d wire. The Fock term couples XY fields at different $\v k$'s, which is identified as a Josephson coupling. In Appendix~\ref{app:SCring} we show that this coupling is local in $k$-space. Crucially, the Berry gauge connection of the continuum band enters the Fock term through the form factor, so that this $k$-space Josephson coupling is gauge invariant.
\begin{align}
    H_{Fock}
    = -\sum_{\v k, \v q} \tilde{V}(\v q, \v k) &\Psi^*_{\v G}(\v k +\v q) \Psi_{\v G}(\v k) \nonumber\\
    & e^{i\int_{\v k}^{\v k+\v q} d\v q' \cdot (A(\v q') -A(\v q'+\v G))} 
\end{align}
where $\tilde{V}(\v q,\v k)$ is $\v k$-dependent because it has absorbed the magnitude of form factor.

As a result, we identify this model as essentially a narrow superconducting ring under a background magnetic field. Note that the effective gauge field is defined by the non-interacting band, and is therefore not dynamical. This situation is analogous to the Little-Parks experiment where a superconducting thin cylinder is pierced by a background magnetic field, such that the winding of the superconducting order parameter is given by the rounding of the piercing flux.

Finally, the effective model defined on the Hilbert space of MBZ corner can be extracted by integrating out the fluctuations in ``superconducting" wires connecting them (see Fig.~\ref{fig:SCring}(b)). We leave the detailed derivation of this model to Appendix~\ref{app:SCring} and simply state the resulting effective Hamiltonian
\begin{equation}
    H_{Fock}[C] = \tilde{J} \left(C-3m-\frac{\Phi_{BZ}}{2\pi}\right)^2
    \label{Eresult}
\end{equation}
where $m$ always takes the integer value that minimizes $E_{Fock}$, which makes the Hamiltonian periodic under $C\rightarrow C+3$. This shows that the Chern number is always rounded to the closest integer to $\frac{\Phi_{BZ}}{2\pi}$, and also predicts the analogous Little-Parks oscillation when $\Phi_{BZ}$ is tuned. Note that Eq.~\ref{Eresult} is consistent with the $C$-dependence predicted by the phenomenological model in Eq.~\ref{GL2}.

\section{Competing Fermi liquid and Hall Crystal Phases}
\label{sec_competing}
In the absence of the \moire potential, a natural candidate for the ground state (in addition to the aforementioned Wigner/Hall crystal states) is the Fermi liquid, which preserves translational symmetry.
To provide a quantitative comparison of the crystal and liquid states in the moir\'e-less setting, we examine their competition within the framework of Hartree-Fock theory.
%In this section, we discuss the competition between the Fermi liquid and Hall crystal states.
%We examine the fate of these states in the absence of the \moire potential.
To that end, the Hall crystal is signified by the breaking of the translational symmetry by selecting (in momentum space) a reciprocal lattice vector $\v{G}$.
As a standard, we take the reciprocal lattice vector to correspond to that of a \moire unit cell with hBN aligned at the experimentally relevant twist angle $0.77^{\circ}$.
In the framework of mean-field theory, this corresponds to $\langle c^{\dag}_{\alpha,\bfk+\textbf{G}} c_{\beta,\bfk} \rangle \neq 0$, where the Greek indices indicate a generalized spin/valley/band degree of freedom.
The Fermi liquid, on the other hand,  preserves continuous translation symmetry; this corresponds to the mean-field density matrix $\langle c^{\dag}_{\alpha,\bfk} c_{\beta,\bfk} \rangle \neq 0$.

%We present in Fig.~\ref{fig_fl_hc_comp}, the Hartee-Fock energy difference between the Fermi liquid and the Hall crystal states.
At the level of Hartree-Fock, the Fermi liquid state is found to be higher in energy than Hall crystal by $\sim 5$meV per unit cell (see Fig.~\ref{fig_fl_hc_comp} in Appendix~\ref{app_fl_hc_comparison}).
This may naively suggest that the moir\'e-less setting is partial to the formation of the electronic crystal phase.
However, it is important to emphasize that the above quantitative comparison is within the framework of Hartree-Fock theory,  and it is important to re-evaluate the FL-AHC competition beyond Hartree-Fock as we elaborate below. 
%Indeed, just as in jellium, a more refined numerical calculation is required to quantitatively determine the true victor of this energy battle. 

\section{Beyond Hartree-Fock: comments on the true phase diagram, and on phenomenology}
\label{sec_beyond}
In jellium, Hartree-Fock theory famously severely overestimates\cite{trail2003unrestricted,bernu2011hartree} the stability of the Wigner Crystal. 
It predicts that the WC crystal forms at an $r_s \approx 1.2 - 1.4$ ($r_s = \frac{l}{a_0}$ is the ratio of the inter-electron spacing $l$ with the Bohr radius $a_0$, and is the standard measure of the ratio of kinetic and Coulomb energies for quadratically dispersing bands). 
Detailed Monte Carlo  calculations\cite{tanatar1989ground,drummond2009phase}, however, show that the true transition to the WC does not occur till an $r_s^{true}  \approx 30$. 
For large values of $r_s < r_s^{true}$, the ground state is a strongly correlated Fermi liquid. 
This state is stabilized relative to the WC by a very small energy\cite{drummond2009phase} (about a fraction of $\mathcal{O}(10^{-3})$  over a large range of $r_s$). 
This tiny stabilization energy can be rationalized by noting that, at short distances and short times, the liquid behaves essentially the same as the solid and thus the two states have the same potential energy. At late times, however, the liquid can flow, and this presumably leads to a slight lowering of the kinetic energy compared to the crystal. 
This observation underlies a picture of the strongly correlated Fermi liquid as an ``almost localized'  or ``nearly frozen" liquid\cite{vollhardt1984normal,castaing1979phase}, as appreciated a long time ago for ordinary liquids\cite{frenkel1946}. 

\begin{figure}[t]
    \centering
    \includegraphics[width=1\linewidth]{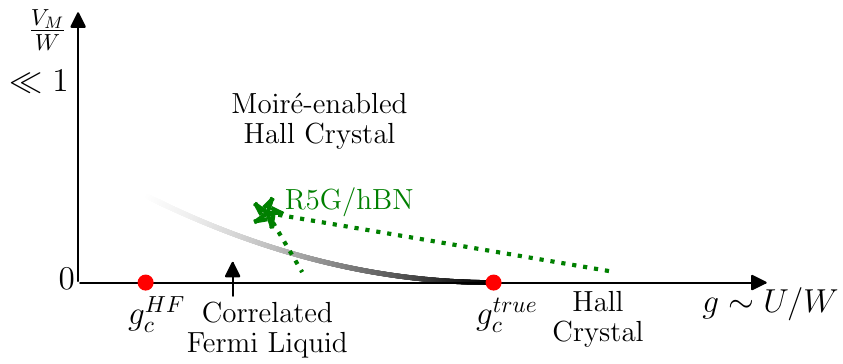}
    \caption{Schematic phase diagram showing the competition between the Fermi liquid and the Hall crystal, both at zero and non-zero \moire potential strength $V_M$. The parameter $g$ is a measure of the ratio of Coulomb ($U$) to kinetic ($W$) energies. Even if the Hall crystal is not the true ground state (beyond Hartree-Fock), it can be stabilized by a weak \moire potential, a state we denote the moir\'e-enabled Hall crystal'.
    A possible location of R5G/hBN is indicated by the green star at a not-too-high displacement field.
    The dashed green lines indicate the possible fate of the system under increasing displacement field. The evolution may either move the system into the pristine AHC phase $g > g_c^{true}$ weakly pinned by the \moire potential, or fall short and land in the FL phase with $g < g_c^{true}$. 
    }
    \label{fig_FLHCVM}
\end{figure}

{\bf Moir\'e-enabled AHC}: 
Returning to R5G, we expect a similar situation. 
Consider the phase diagram at a density where the Hall crystal is stabilized in the Hartree-Fock approximation.  
Let $g$ be a dimensionless parameter that measures the ratio of Coulomb ($U$) and kinetic ($W$) energies. The Coulomb scale is $U = \frac{e^2}{\epsilon l}$ where $\epsilon$ is the dielectric constant, and $l$ is the inter-electron spacing. The scale $W$ for the kinetic energy can be taken to be the bandwidth within the first MBZ with the \moire potential turned off. 
Then we expect that the true $g_c^{true}$ at which the transition\footnote{In principle, just like in discussions of jellium (see Ref. \onlinecite{kim2023dynamical} for a recent discussion), intermediate phases between the Fermi liquid and crystal may appear, particularly with unscreened long-ranged Coulomb interactions\cite{spivak2004phases} but do not affect the essential point we make below.} to the Hall crystal occurs to be (substantively) larger than the Hartree-Fock $g_c^{HF}$. 
This is depicted schematically in Fig. \ref{fig_FLHCVM}. For $g_c^{HF} < g < g_c^{true}$, we expect that the ground state is a correlated Fermi liquid that is `almost localized'. 
Further, we expect that the ground state energy per particle of the Fermi liquid is smaller by only a small amount compared to that of the Hall crystal. 

We leave the precise phase diagram as an interesting target for future numerical work using, say, Variational Monte Carlo methods. 
For now, consider the phase diagram in the presence of a periodic \moire potential. 
The expected small stabilization energy of the Fermi liquid over the Hall crystal implies that even a weak \moire potential will induce a transition from Fermi liquid to a Chern insulator. 
This is because the Hall crystal gains commensuration energy in the \moire potential which will overcome the small energy by which the Fermi liquid is stabilized. 
To see this, consider the energy (per particle) of both the Fermi liquid and the Hall crystal. 
We denote these $E_{FL}(g, V_M)$ and $E_{HC}(g, V_M)$ respectively. At the (first order) FL-HC transition at $V_M = 0$, we have 
\begin{equation} 
  E_{FL}( VM  = 0, g_c^{true}) = E_{HC}( V_M = 0, g_c^{true}).
  \end{equation} 
  As $g$ decreases slightly below $g_c$, we will have 
\begin{equation}  
E_{FL}( V_M  = 0, g) = E_{HC}( V_M = 0, g) - a(g_c^{true} - g) + \cdots 
\end{equation} 
with $a > 0$, and the ellipses represent higher-order terms in $(g_c^{true} - g)$. 
Now let us consider the energies when a small $V_M$ is turned on. The crystal will lower its energy at linear order in $V_M$: 
 \begin{equation}
  E_{HC}( V_M, g) = E_{HC}( V_M = 0, g) - b V_M + \cdots 
\end{equation} 
where now the ellipses represent higher order terms in $V_M$, and $b >0$. However the Fermi liquid will only have a change at quadratic order in $V_M$: 
\begin{equation} 
E_{FL}( V_M , g) = E_{FL}( V_M  = 0, g) - c (V_M )^2 + \cdots 
\end{equation} 
with $c > 0$. Thus the Hall crystal will win over the Fermi liquid when 
\begin{equation} 
b V_M - c V_M^2 > a (g_c^{true} - g) 
\end{equation} 
It follows that the phase boundary will have the shape shown in Fig.~\ref{fig_FLHCVM}. (For a recent numerical study\footnote{We thank Ilya Esterlis for discussions on this paper and related questions} of a similar phase competition between the FL and the ordinary WC in TMD moire materials, see Ref. \cite{yang2023metal}). 
Further, we expect that even somewhat far below $g_c^{true}$, the critical $V_M$ needed to stabilize the Hall crystal is small. The \moire induced FL to (pinned) Hall crystal state will be first order (at least with a screened Coulomb interaction). 

We call the Chern insulating state induced by $V_M$ a `moir\'e-enabled AHC'. This state is likely stable over a much wider range of the phase diagram than the pristine AHC which exists in the strictly translation-invariant system.

We do not of course know where R5G/hBN sits in this phase diagram, but it is conceivable that in a range of displacement fields, it is in the moir\'e-enabled AHC regime (denoted by green star in Fig.~\ref{fig_FLHCVM}). If that is the case, the moir\'e-less ({\em i.e.} unaligned $R5G$) will be a Fermi liquid metal (with spin/valley polarization). 
Aligning with hBN will, in a range of twist angles, push the system across the phase boundary to the moir\'e-enabled AHC. 
So long as the effective $V_M$ felt by the electrons in the occupied layers is small (as is expected at the large displacement fields needed to stabilize the QAH states in $R5G$), the charge gap of the moir\'e-enabled AHC state will be determined by the strength of the Coulomb interaction, and not by the strength of the \moire potential. Thus, even if the \moire is needed to stabilize the Chern insulator, the physics is still primarily determined by the Coulomb interactions. The sole role of the \moire potential is to tip the delicate balance of energy between the Fermi liquid and Hall crystal. 

Note that increasing the displacement field has a number of different effects: it changes the ratio of kinetic and Coulomb energies and thus increases $g$; it also reduces the effective $V_M$ as the conduction electrons get driven further away from the aligned hBN side; finally, it also changes the Berry curvature distribution. Changing Berry curvature is not included in the schematic phase diagram Fig.~\ref{fig_FLHCVM}; we have already discussed its role in determining the selection between the AHC and the ordinary WC. In the context of Fig.~\ref{fig_FLHCVM}, increasing $|u_d|$ thus corresponds to both increasing $g$ and decreasing $V_M$. which thus pushes the system closer to the pristine AHC, but may also take it back to the FL. 

{\bf Doping away from commensuration}: 
It is also very interesting to ask what happens at densities that are not commensurate with the \moire potential, assuming we are at $g < g_c^{true}$. For instance, consider decreasing the density so that if any crystal forms it will have a larger lattice spacing than the \moire potential. If the density deviation from the commensurate value is small, we may expect that the crystal pays the cost of the extra elastic energy to lock its period to the \moire lattice so as to gain commensuration energy. The density deficit will then be accommodated by a non-zero density of vacancies (equivalent, in a weak coupling picture, to `doping the Chern band'). Such a state may still overcome the Fermi liquid if the commensuration energy is big enough. These vacancies (i.e the doped holes) can then form FQAH states at the suitable fillings. Alternately, if $g$ is not close enough to $g_c^{true}$, the Fermi liquid may win out at the lower density and the system forms a metallic state. The locking to the \moire potential becomes less likely the lower the density is. Thus at low densities (well below $\nu = 1$) the primary competition will be between the Fermi liquid and a crystal (either AHC or WC) that has its natural intrinsic period. However, from the Hartree-Fock calculations, we also know that at low density ({\em i.e} low $\theta$), the WC is favored over the AHC as the Berry curvature in the first MBZ of the non-interacting band is small. Thus at densities well below $\nu = 1$, the two main players will be the Fermi liquid and the WC with the latter winning at the lowest densities. 
These expectations are qualitatively consistent with the phase diagram reported in  Ref. \onlinecite{lu2023fractional}. 

A more radical possibility is that of a fractional AHC, where even for $g > g_c^{true}$ and in the absence of \moire, the system prefers to lock in a specific density and accommodate any density deficit through vacancies that form an FQAH state. While we are not aware of a strong reason forbidding such a state to exist, it is likely not stable energetically at densities so low that half the lattice is empty. Thus, this possibility may not be supported in the current experiments where FQAH states are seen down to filling $\nu = 2/5$; however it is an interesting possibility that could be relevant as more RnG systems are studied.

{\bf Disorder effects}: 
Finally, we briefly discuss what we might expect in moir\'e-less R5G in the presence of weak disorder. If the ground state in the clean limit is an AHC, then it will be randomly pinned by the impurities (and long-range crystalline order will be lost). The resulting state will be a disordered IQAH insulator. If $g < g_c^{true}$, and the ground state is a correlated Fermi liquid, then locally near each impurity, we might expect, for the same reason as above, that the delicate balance between FL and AHC is tipped in favor of the AHC beyond a non-zero but small value. Thus we expect puddles of randomly oriented AHC to nucleate within the metallic state (ignoring Anderson localization effects which will not set in up to parametrically larger scales for weak disorder). At stronger disorder strength, there will be a transition to the disordered IQAH insulator. 

In transport experiments, the sliding motion of the ideal, clean AHC will lead to an infinite longitudinal conductivity, and a finite Hall conductivity (so that the Hall resistivity is zero). Once the AHC is pinned, either by a periodic potential, or by disorder, the DC longitudinal conductivity at low bias voltage will be zero while the Hall conductivity will be quantized, exactly as expected of an IQAH insulator. In the clean Fermi liquid metallic state,  with very weak disorder, the Berry curvature enclosed within the Fermi surface ($\Phi_B$) will lead to an anomalous Hall conductivity $\sigma_{xy}^{FL} = \frac{e^2}{h} \frac{\Phi_B}{2\pi}$ but this will be much smaller than the longitudinal conductivity $\sigma_{xx}^{FL} = \frac{e^2}{h} \left(K_F l_{mf} \right)$ ($K_F$ is the Fermi momentum and $l_{mf}$ is the mean free path) so that the Hall resistivity will be small. As the disorder increases, there will be the puddles of AHC nucleated by the disorder, which will presumably lead to an enhanced, but not quantized, Hall resistivity.

%\section{Higher Chern bands at larger twist angle and enhanced \moire potential}

\section{Routes to higher Chern bands in electron-doped RnG}
\label{sec:highC}

The discussion of the interaction-induced Chern band/AHC in RnG has primarily focused on fully/partially filling a $|C|=1$ band \cite{dong2023theory,zhou2023fractional,dong2023anomalous}. 
In this section, we examine the possibility of finding (within Hartree-Fock) similar interaction-induced IQAH states with  $|C|>1$ for electron-doped RnG. 
We explore two routes where such a $|C| > 1$ state may occur. In the calculations below, we have included the warping terms that have been ignored so far. 

\begin{figure}[t]
    \centering
    \includegraphics[width=1\linewidth]{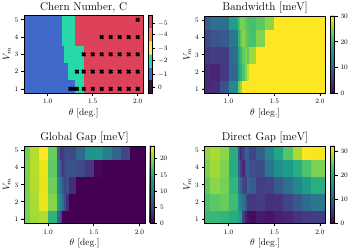}
    \caption{Hartree-Fock bands indicating Chern number and bandwidth of the active band, global band gap, and direct band gap to nearest conduction band for twist angles ($\theta$) and \moire potential strengths $(V_m$) of the full 10-orbital model (including all warping terms).
    The displacement field energy is $u_d = -36$ meV, and the dielectric constant is $\epsilon = 8$.
    The Chern number is well-defined when the direct band gap is non-zero ($\geq0.5$meV).
    The `crossed' out yellow boxes in the Chern number indicate cases where the global bandgap is zero ($<0.5$meV), while the direct band gap is still non-zero (i.e. there is an indirect band gap -- the system is metallic).
    Phase diagrams are for a mesh of 21$\times$21, with 4 conduction bands in total (1 active band and 3 remote bands above), and constructed using 19 \moire Brillouin zones, and generated using at least six distinct initial mean-field ansatzes for a given parameter point.
    }
    \label{fig_hf_tuning_moire}
\end{figure}

Firstly, we consider the Hartree-Fock phase diagram in R5G/hBN as a function of the twist angle $(\theta)$ and the strength  $V_m$ of the \moire potential. Though naively the microscopic \moire potential  is fixed by the alignment to the hBN, there is uncertainty on details like the lattice relaxation that may modify it. Thus we simply take the overall magnitude of $V_M$ as a tuning parameter to illustrate the possibilities. 
We present in Fig.~\ref{fig_hf_tuning_moire} the HF-phase diagram for these \moire tuning parameters, where we depict the Chern number, bandwidth, and gaps of the active band. As seen, for the naive \moire strength, the $|C|=1$ state is the only state that is well isolated; the ‘crossed’ out yellow boxes in the Chern number indicate cases where the global bandgap is zero ($<0.5$ meV), while the direct band gap is still non-zero (i.e. there is an indirect band gap and the system is metallic). However, with an enhanced \moire potential to open up the band gap, we can obtain high Chern number bands that are well isolated.
In these regions, the mean-field bandwidth is $\sim 20$ meV, with a direct gap of $\sim 5 - 10$ meV (and a global gap of $3-6$ meV).
In particular, topologically non-trivial bands of $|C| = \{2,5\}$ are formed.

\begin{figure}[t]
    \centering
    \includegraphics[width=0.6\linewidth]{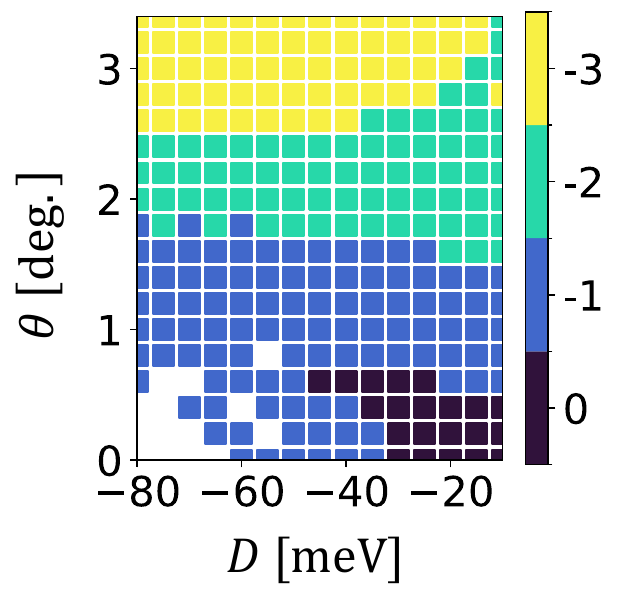}
    \caption{Phase diagrams in $\theta-u_d$ space indicating the Chern number, obtained with the same non-interacting parameters as in Fig.~\ref{hf_pd_compare}, but with an enhanced interaction $\epsilon=2$ (as compared to $\epsilon=6$). This result is obtained for the simplified model without warping terms. We use a momentum mesh of $18\times18$, and include 4 conduction bands.}
    \label{pd_highC}
\end{figure}

Secondly, we examine the regime of strong interactions (enabled by imagining decreasing the dielectric constant $\epsilon$) without warping terms.
We present in Fig.~\ref{pd_highC} the possible Chern numbers as a function of twist angle and displacement field energy for enhanced electron-electron interactions.
Once again, in addition to $|C|=1$ state, higher Chern bands are found to develop, for increasing twist angles.

These routes that lead to bands with enhanced Chern number provide the intriguing possibility to realize incompressible states by fully/partially filling these higher-Chern bands. We hope that these results provide enough motivation for future experimental studies of the higher twist angle region, and more detailed theoretical modeling. 
%Engeneering \moire potentials (need to cite Cano et al papers) would be a route to the development such higher-Chern bands.

\section{Discussion}
\label{sec_discussion}
In this work, we examined the origins of the interaction-induced IQAH in R5G/hBN, and the closely related AHC in \moire-less R5G. 
Focussing on an effective continuum model of R5G, we trace the origin of $|C|=1$ band to the appreciable flux contained about the zone center and the crucial but differing roles played by the Hartree and Fock interaction terms. At a slightly larger alignment angle (or equivalently charge densities corresponding to lattice filling $\nu = 1$) than in current experiments, the Fock term alone is enough to both open a band gap and to endow the occupied band with Chern number. At lower alignment angles, including the one in current devices, the Fock term opens the gap while the preference for the non-zero Chern number comes from the Hartree term. 

We reiterate that, though the `dangling edge' model of pentalayer graphene (effective $k^5$-two-orbital model) captures the ring-like feature of the continuum Berry curvature, it importantly misses on the approximately $-\pi$-Berry flux contained about $k=0$, that becomes ever-increasingly important at large displacement fields.

Examination of the continuum model, in the absence of trigonal warping terms (that are included in quantitative studies of multilayer graphene) and the \moire potential, provides a remarkably intuitive understanding of the Hartree-Fock results.
For large ($0.9^\circ<\theta<1.5^\circ$) twist angles, where the MBZ boundary intersects the Berry-curvature ring, the MBZ contains appreciable Berry flux (between $\pi$ and 3$\pi$), leading the Fock term to merely introduce band gaps and shunt the subsequent flux to the closest integer.
The rounding of Berry curvature is analogous to the vortex quantization in the Little-Parks experiment, as is made explicit by the ``superconducting ring'' model.

At the intermediate twist angle regime, relevant for the existing experiments \cite{lu2023fractional},  the importance of the Hartree interaction in developing a $|C|=1$ band is understood heuristically in terms of the topologically non-trivial band possessing more spread-out charge distribution. Our simplified model provides a quantitative justification for this conclusion. 

The ultimate fate of the non-\moire and the \moire R5G relies on the competition of correlated Fermi liquid, Wigner crystal, and anomalous Hall crystal states. In addition to Hartree-Fock estimates of the energy differences, we discussed simple expectations for what may happen beyond Hartree-Fock. Specifically, even if a correlated FL, rather than the AHC,  is the true ground state in moir\'e-less RnG,  even a very weak \moire potential can tip the Hall crystal to be lower in energy over the Fermi liquid state. This is because we anticipate that, in the moir\'e-less setting (by analogy with the usual two-dimensional electron fluid),  the FL and AHC will likely have very similar ground state energy, and the AHC can gain commensuration energy from the periodic potential. This simple observation allows us to envision the possibility that such a moir\'e-enabled AHC (if not the pristine AHC itself) occupies a wide region of the phase diagram (at the fixed density corresponding to $\nu = 1$, as a function of interaction strength and \moire potential strength), even beyond Hartree-Fock. 
It is clearly important to study these questions with robust numerical methods (such as Variational Monte Carlo)  in the future. 

 Finally, by exploring the phase diagram at slightly larger alignment angles than in current devices, we suggest that it may be possible to stabilize interaction-induced higher Chern bands/AHCs in electron-doped RnG, which too will be interesting to study further in the future.

While this paper was being written, three interesting papers that overlap with ours - Ref. \onlinecite{zeng2024sublattice, tan2024parent, soejima2024anomalous} - appeared. Ref. \onlinecite{zeng2024sublattice} studies the competition between AHC and WC through a pseudospin model defined in k-space.  Ref. \onlinecite{tan2024parent} studies a toy model with constant Berry curvature where the Fock term alone stabilizes an AHC. Ref. \onlinecite{soejima2024anomalous} considers the specific setting appropriate to RnG, and introduces a hot-spot model for the MBZ corners to discuss the stability of the AHC with some rough similarities to our discussion.

\acknowledgements
We thank Ray Ashoori, Ilya Esterlis, Tonghang Han, Long Ju, Kyung-Su Kim, Steve Kivelson, Zhengguang Lu, Ashvin Vishwanath, Mike Zaletel and particularly Boran Zhou and Ya-Hui Zhang for many inspiring discussions. We are especially grateful to Tomo Soejima and Junkai Dong for illuminating discussions. TS was supported by NSF grant DMR-2206305, and partially through a Simons Investigator Award from the Simons Foundation. This work was also partly supported by the Simons Collaboration on Ultra-Quantum Matter, which is a grant from the Simons Foundation (Grant No. 651446, T.S.). 
The authors acknowledge the MIT SuperCloud and Lincoln Laboratory Supercomputing Center for providing HPC resources that have contributed to the research results reported within this manuscript.

\maketitle
\bibliographystyle{apsrev4-1}
\bibliography{FQAH}

\newpage

\onecolumngrid

\appendix

\section{Full-10 band tight-binding model for pentalayer graphene}
\label{app_ten_band_model}

For clarity, we present the full-10 band tight-binding model (for a given valley $K/K'$ and spin $\uparrow/\downarrow$) for pentalayer graphene (see also Ref. \cite{dong2023theory}):

%%
%\begin{widetext}
\begin{align}
\label{ten_band_model_penta}
H = 
\begin{pmatrix}
2 u_d & v_0^{\dag} & v_4^{\dag} & v_3 & 0 & \frac{\gamma_2}{2} & 0 & 0 & 0 & 0\\
v_0 & 2 u_d + \delta & \gamma_1 & v_4^{\dag} & 0 & 0 & 0 & 0 & 0 & 0 \\
v_4 & \gamma_1 & u_d + u_a & v_0^\dag & v_4^\dag & v_3 & 0 & \frac{\gamma_2}{2} & 0 & 0 \\
v_3^{\dag} & v_4 & v_0 & u_d + u_a & \gamma_1 & v_4^{\dag} & 0 & 0 & 0 & 0 \\
0 & 0 & v_4 & \gamma_1 & u_a & v_0^{\dag} & v_4^{\dag} & v_3 & 0 & \frac{\gamma_2}{2} \\
\frac{\gamma_2}{2} & 0 & v_3^{\dag} & v_4  & v_0  & u_a & \gamma_1 & v_4 ^{\dag} & 0 & 0 \\
0 & 0 & 0 & 0 & v_4 & \gamma_1 & -u_d + u_a & v_0^{\dag} & v_4^{\dag} & v_3 \\
0 &0 & \frac{\gamma_2}{2} & 0 & v_3^{\dag} & v_4 & v_0 & -u_d + u_a & \gamma_1 & v_4^{\dag} \\
0 & 0 & 0 & 0 & 0 & 0 & v_4 & \gamma_1 & -2u_d + \delta & v_0^{\dag}\\
0 & 0 & 0 & 0 & \frac{\gamma_2}{2} & 0 & v_3^{\dag} & v_4 & v_0 & -2u_d  
\end{pmatrix},
\end{align}
%\end{widetext}
%%
where we employ the basis of $(A_1, B_1, A_2, B_2, A_3, B_3, A_4, B_4, A_5, B_5)$.
We have also used the notation of $v_i = \frac{\sqrt{3}}{2} t _l (\pm k_x + i k_y)$ for the monolayer graphene form factors, where $k_{x,y}$ are small momenta expanded about the valley of interest, and the layer is denoted by $l$.
Note that $\gamma_l \equiv \frac{\sqrt{3}}{2} t_l$.
In the main text, $\gamma_{1,2,3,4}$ are referred to as `warping terms', which correspond to the trigonal warping of rhombohedral graphene.
We direct the reader to Ref. \onlinecite{{dong2023theory}} for the model parameter values.

\section{Density profile of Bloch function at high symmetry momentum}
\label{app:density_profile}
In the following, we formulate a theory by establishing connections between the symmetry indices at the high-symmetry momenta and the density profile of the corresponding Bloch functions.

To begin with, the eigenstate for the mean-field Hamiltonian at the $\v K_M$ point is
\be
    \psi_{\v K_M}(\v r) = \sum_{i=1,2,3} \alpha_i \psi_{\v K_i}(\v r)
\ee
where $\v K_M$ is the \moire $K$ point measured from the graphene Dirac point $K_D$ corresponding to valley $+$, and $K_i$'s are the three equivalent \moire $K$ points. 
The 10-component wavefunction $\psi_{\v k}(\v r)=e^{i(\v k+\v K_D)\v r}u^0_{\v k}(\v r)$ is the plane wave corresponding to the lowest conduction band (denoted by $^0$) of the continuum model. Expressed in the 10-orbital basis described in Sec.~\ref{sec_continuum_model_begin}, the periodic Bloch function is $u^0_{\v k}(\v r)=(u_{A_1\v k}, u_{B_1\v k}, u_{A_2\v k}, u_{B_2\v k}, u_{A_3\v k}, u_{B_3\v k}, u_{A_4\v k}, u_{B_4\v k}, u_{A_5\v k}, u_{B_5\v k})^T$.
Alternatively, one can express the Bloch function $u^0$ in real space, instead of the 10-orbital basis,
\be
    u^0_{\v k}(\v r) = \sum_{n} u^0_{n,\v k}(\v r) = \sum_{n,\v R} u_{n}(\v k) \delta^2(\v r-\v r_n-\v R),
\ee
where $\v r$ is a 3-dimensional position vector, $\v R$ labels the R5G unit cell, and $\v r_n$ is the displacement of the $n^{\text{th}}$ of the orbital (in the 10-orbital basis) within the R5G unit cell.

The prominent $C_3$ axis is defined to run through a $B_5$ site at $r=0$, and its corresponding operation is, % The operation acts as
\be
    \v r \rightarrow C_3 \circ \v r .
\ee
This operation transforms $\psi_{\v K_M}(\v r)$ into
\begin{align}
\label{1}
    C_3 \circ \psi_{\v K_M}(\v r) &= \psi_{\v K_M}(C_3 \circ \v r) = \sum_{i=1,2,3} \alpha_i \psi_{\v K_i}(C_3 \circ \v r).
\end{align}
We note that there is a gauge degree of freedom $\psi_{\v k}\rightarrow \psi_{\v k}e^{i\theta_{\v k}}$. 
To fix a gauge, we demand that the plane wave solution for the continuum model preserves $C_3$, explicitly
\be
\label{gaugechoice}
    \psi_{\v k}(C_3 \circ \v r) = \psi_{C^{-1}_3 \circ \v k}(\v r) ,
\ee
so that there is no ambiguity for $\v k=\v \Gamma$. With this gauge choice, from Eq.~\ref{1} we immediately get
\be
    C_3 \circ \psi_{\v K_M}(\v r) = \sum_{i=1,2,3} \alpha_i \psi_{\v K_i}(C_3 \circ \v r) = \sum_{i=1,2,3} \alpha_{i+1} \psi_{\v K_i}(\v r)
    \label{eq_c3_operation}
\ee
where we have made the convention $\v K_{i+1} = C_3 \circ \v K_i$. 
The conclusion from Eq.~\ref{eq_c3_operation} is striking: the $C_3$ operation cyclically permutes the coefficients $(\alpha_1, \alpha_2, \alpha_3)$. 
The $C_3$ permutation has an eigenvalue of $-1, 0, 1$.

The eigenvalues of the $C_3$ operation provide the location of the peak of the Bloch function (of $\v K_M$) in the \moire unit cell.
%In the following, we want to demonstrate that the eigenvalues indicate where the Bloch function of $\v K_M$ peaks in the moire unit cell. 
For this purpose, we separate out the fastly-varying momentum $K_D$ from the global phase,
%$e^{i\v K_D\v r}$,
\be
    \psi_{\v K_M} = \sum_{i} \alpha_i \psi_{\v K_i}(\v r) =e^{i\v K_D\v r} \sum_{i} \alpha_i e^{i\v K_i \v r}u^0_{\v K_i}(\v r),
\ee
to obtain the amplitude on orbital $n$,
\be
    \left |\psi^n_{\v K_M}(\v r)\right | = \left | \sum_{i} \alpha_i e^{i\v K_i \v r}u_{n\v K_i} \right |.
\ee
From the gauge choice of Eq.~\ref{gaugechoice}, the $C_3$ transformation on the periodic part of the Bloch function ($u_{n \v K_i}$ for $i=1,2,3$) generates a relative phase,
%To determine the peak location, the remaining work is to figure out the relative phase of $u_{n \v K_i}$ for $i=1,2,3$, which should be determined by our earlier gauge choice Eq.~\ref{gaugechoice}.

\begin{align}
    C_3 \circ \psi_{\v K_i}(\v r)
    &= e^{i (\v K_i+\v K_D) \cdot (C_3 \circ \v r)}u^0_{\v K_i}(C_3 \circ \v r)\nonumber \\
    &= e^{i C_3^{-1} \circ (\v K_i+\v K_D) \cdot \v r}u^0_{\v K_i}(C_3 \circ \v r)\nonumber \\
    &= e^{i (\v K_{i-1}+\v K_D) \cdot \v r} \left(e^{i\v G_g \cdot \v r}u^0_{\v K_{i}}(\v r)\right)\label{2}\\
    &\equiv \psi_{C_3^{-1} \circ  \v K_i}(\v r)\nonumber\\
    &= e^{i (\v K_{i-1}+\v K_D) \cdot \v r} u^0_{\v K_{i-1}}(\v r) \label{3}.
\end{align}
To obtain Eq.~\ref{2}, we have used the periodic nature of $u^0_{\v K_i}(\v r)$, and $C_3 \circ \v r = \v r (\text{mod} \; \v R)$, and $C_3^{-1} \circ \v K_D = \v K_D + \v G_g$, where $\v G_g$ is a reciprocal lattice vector for the graphene lattice.  Then it becomes clear that the bracketed term in Eq.~\ref{2} is,
\begin{align}
\label{4}
    e^{i\v G_g \cdot \v r}u^0_{\v K_{i}}(\v r) &= e^{i \v G_g \cdot \v r} \sum_{n,\v R} u_{n\v K_{i}} \delta^2(\v r-\v r_n-\v R)\nonumber \\
    &=  \sum_{n, \v R} e^{i \v G_g \cdot \v r_n} u_{n\v K_{i}} \delta^2(\v r- \v r_n - \v R),
\end{align}
where we employ the delta-function constraint.
Comparing Eq.~\ref{2} with \ref{3}, and combining with Eq.~\ref{4}, we find the relation between the 10-component Bloch state vector $u_{n\v k}$ to its $C_3$ transformed counterpart ($u_{n (C_3 \circ \v k)}$),
\be
    u_{n(C_3 \circ \v k)} = S_{nn} u_{n\v k}
\ee
where $S$ is a diagonal matrix $S=e^{-i \v G_g \cdot \v r_n} = \text{diag}(\omega^2, \omega^1, \omega^1, 1, 1, \omega^2, \omega^2, \omega^1, \omega^1, 1)$, with $\omega=e^{i2\pi/3}$. 
This leads to,
\be
     u_{n\v K_{i+1}} = S_{nn} u_{n\v K_i}.
\ee
Thus, for a Bloch state with $C_3$ symmetry index $L$, we have the corresponding eigenvalues $\alpha=(1, \omega^L, \omega^{2L})$. 
The corresponding amplitude is,
\be
    \rho^n_{\v K_M}(r) = \left |\psi^n_{\v K_M}(\v r)\right | ^2 \sim \left | \sum_{j=1,2,3} \omega^{jL} S_{nn}^{j} e^{i\v K_j \v r} \right | |u_{n\v K_1}|^2.
\ee
The peak location is where the three plane waves constructively interfere.

\begin{figure}[t]
    \centering
    \includegraphics[width=0.5\linewidth]{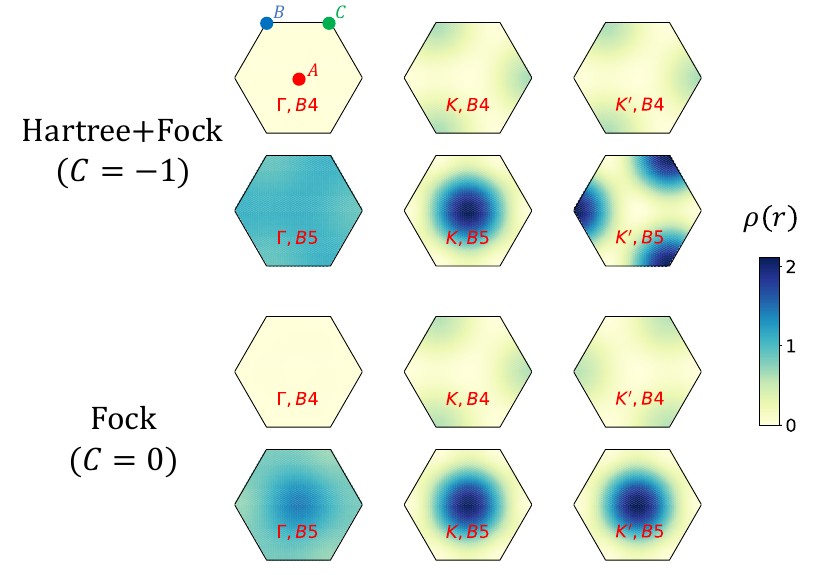}
    \caption{Real-space $B_4$ and $B_5$ density profile for Bloch functions at high symmetry momentum. 
    With the Hartree term included, the $B_5$ orbitals repel and the density profile is spread out to different regions of the moir\'e unit cell, while the $B_4$ orbitals overlap. 
    Without the Hartree term, the Fock energy is reduced by overlapping $B_5$ orbitals. 
    These results are obtained for $\theta=0.7^\circ$, $u_d=-30$meV, and with a very weak moire potential $C_{AA}=C_{BB}=1$meV to pin the wavefunction to high symmetry positions in the unit cell.}
    \label{denisty_profile}
\end{figure}

For example, for a state with $C_3$ index $L=0$, the density profile on orbital $B_5$ (meaning $S_{nn}=1$) peaks at origin $\v r=0$, while for $L=1$, the peak of $B_5$ density shifts to a corner of the hexagonal \moire unit cell $\v r_1$, which satisfies $e^{i\v K_j \v r_1} = \omega^{-j}$, so that the j-dependent phase $e^{jL}$ gets canceled at $\v r_1$.
Similarly, the peak of density for orbital $B_5$ and $B_4$ are located at different high symmetry locations in the hexagonal \moire unit cell (see Fig.~\ref{denisty_profile})

Finally, note that $\v K_M$ and $\bar{\v K}_M$ are connected through a $C_2$ rotation. 
As their symmetry indices interpolate from $L=0$ to $L=1$, the peaks for $\rho^n_{\v K_M}(r)$ and $\rho^n_{\bar{\v K}_M}(r)$ get shifted from the origin to opposite corners of the hexagonal \moire unit cell. 
As a result, nontrivial $C_3$ indices at $\v K_M$ and $\bar{\v K}_M$ would reduce the Hartree energy between these two Bloch states, due to minimizing the wavefunction overlap.
This is why the Hartree term selects the set of symmetry indices to be $(L_\Gamma, L_{\v K}, L_{\bar{\v K}}) = (0, 1, 1)$, which gives $C=-1$.

\section{Effect of Translation on the $C_3$ indices }
\label{symmetry}
\begin{figure}[h]
    \centering
    \includegraphics[width=0.2\linewidth]{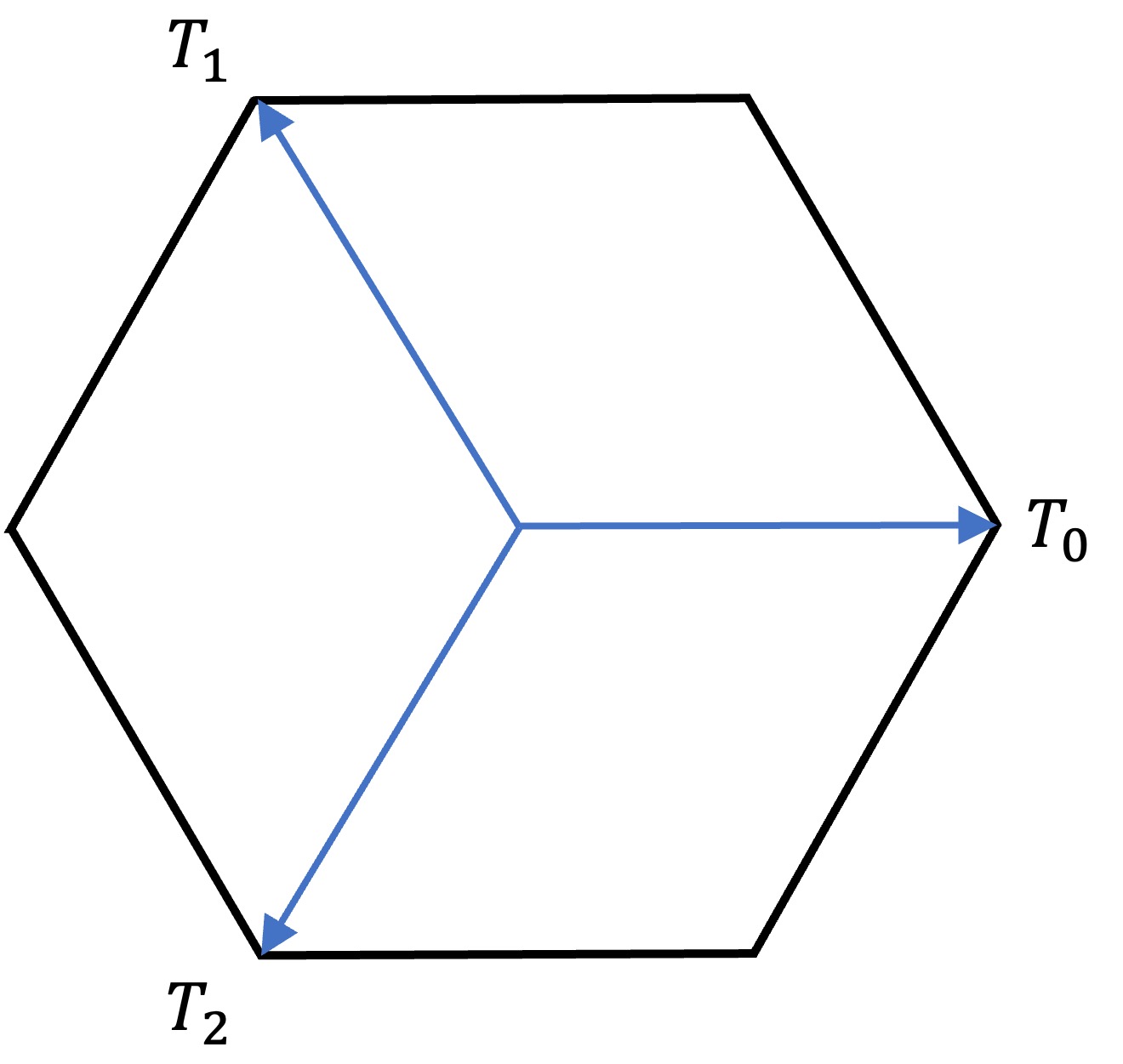}
    \caption{Definition for translation vectors.}
    \label{fig_translation}
\end{figure}

In this appendix, we explain how the translation operation transforms the $C_3$ indices at $K_M$ and $\bar{K}_M$ points. This data is useful in containing the possible terms in the simplified model for competing crystalline states (see Appendix~\ref{app_alternative_gl_model}). Under $C_3$ rotation, the wavefunction for moir\'e band transforms as
\begin{align}
    C_3 \ket{\psi_{K_M}} = \omega^l \ket{\psi_{K_M}}\\
    C_3 \ket{\psi_{\bar{K}_M}} = \omega^s \ket{\psi_{\bar{K}_M}}
\end{align}
where $l$ and $s$ are the $C_3$ indices for $K_M$ and $\bar{K}_M$. Under translation,
\begin{align}
    T_i \ket{\psi_{K_M}} = \omega^i \ket{\psi_{K_M}}\\
    T_i \ket{\psi_{\bar{K}_M}} = \omega^{-i} \ket{\psi_{\bar{K}_M}}
\end{align}
where $T_{i+3}=T_i$, $C_3 T_i C_3^{-1} = T_{i+1}$, $i=0,1,2$ corresponds to the three equivalent corners of the \moire unit cell (see Fig.\ref{fig_translation}).
Note
\be
    C_3 T_i = C_3 T_i C_3^{-1} C_3 = T_{i+1} C_3
\ee
Therefore,
\be
    C_3 (T_i \ket{\psi_{K_M}}) = T_{i+1} C_3 \ket{\psi_{K_M}} = \omega^{l+i+1}\ket{\psi_{K_M}} = \omega^{l+1} (T_i \ket{\psi_{K_M}})
\ee
\be
    C_3 (T_i \ket{\psi_{\bar{K}_M}}) = T_{i+1} C_3 \ket{\psi_{\bar{K}_M}} = \omega^{s-i-1}\ket{\psi_{\bar{K}_M}} = \omega^{s-1} (T_i \ket{\psi_{\bar{K}_M}})
\ee
We have seen that the translations transform the $C_3$ indices at $(K_M, \bar{K}_M)$ into
\be
    T_i: (l, s) \rightarrow (l+1, s-1)
\ee

\begin{figure}[h]
    \centering
    \includegraphics[width=0.2\linewidth]{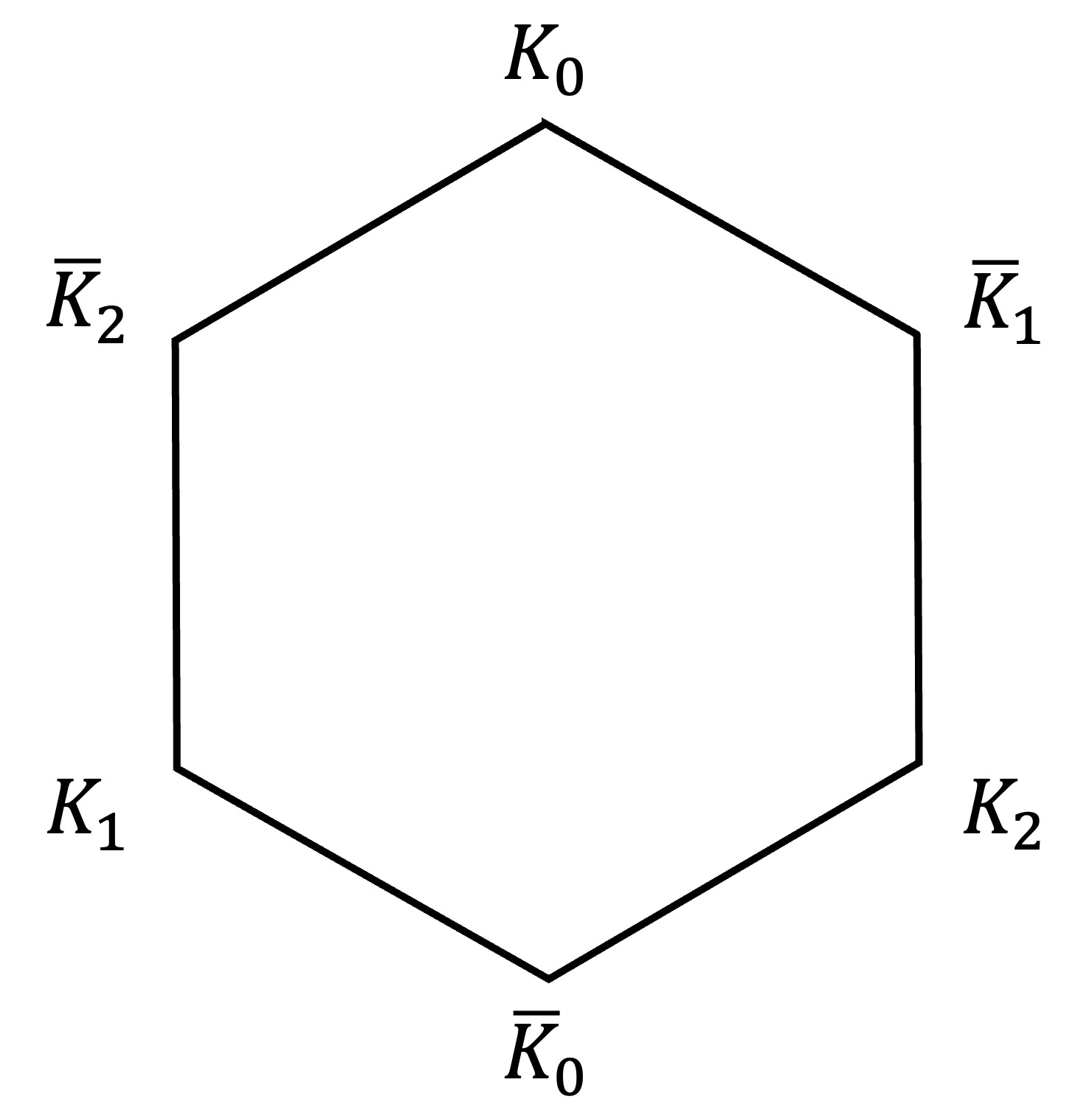}
    \caption{Labeling for MBZ corners.}
    \label{fig:BZ}
\end{figure}

\section{Pseudopotential in angular momentum channels}
\label{app_alternative_gl_model}

Continuing from Sec.~\ref{sec_pseudopotential} in the main text, in this section we detail the derivation of constants $J$ and $\phi$ in the simplified model Eq.~\ref{GL2}. We first provide a deliberately careful derivation. For readers who favor a more compact derivation, we direct them to start from Eq.~\ref{compact}.

We begin with the interacting term of Hamiltonian projected to the lowest conduction band of the continuum model
\be
    H= \sum_{\v q,\v k,\v k'} V_{\v q} \lambda_{\v k, \v k+ \v q} \lambda_{\v k',\v k'-\v q} c^\dagger_{\v k} c^\dagger_{\v k'}c_{\v k'-\v q}c_{\v k+\v q}
\ee
where $\lambda_{\v k, \v k'}=\langle u_{\v k}|u_{\v k'}\rangle$ is the form factor.
We focus on the terms within the Hilbert space of two MBZ corners $K_M$ and $\bar{K}_M$. In the following, we will denote the two sets of equivalent momentum at two corners of BZ as $K_i$ and $\bar{K}_i$, where $i=0,1,2$ and $K_{i+1}=C_3 K_i$, $K_{n+3}=K_{n}$.
\begin{align}
    H
    = \sum_{nn'mm'} &V_{K_n-K_{n'}} \lambda_{K_n,K_{n'}} \lambda_{K_m,K_{m'}} \delta^2(K_n-K_{n'}+K_m-K_{m'}) 
 c^\dagger_{K_n}c^\dagger_{K_m}c_{K_{m'}}c_{K_{n'}}\nonumber\\
    + &V_{K_n-K_{n'}} \lambda_{K_n,K_{n'}} \lambda_{\bar{K}_m,\bar{K}_{m'}} \delta^2(K_n-K_{n'}+\bar{K}_m-\bar{K}_{m'}) c^\dagger_{K_n}c^\dagger_{\bar{K}_m}c_{\bar{K}_{m'}}c_{K_{n'}}\nonumber\\
    + &V_{K_n-\bar{K}_{n'}} \lambda_{K_n,\bar{K}_{n'}} \lambda_{\bar{K}_m,K_{m'}} \delta^2(K_n-\bar{K}_{n'}+\bar{K}_m-K_{m'}) c^\dagger_{K_n}c^\dagger_{\bar{K}_m}c_{K_{m'}}c_{\bar{K}_{n'}}\nonumber\\
    + &V_{K_n-\bar{K}_{n'}} \lambda_{K_n,\bar{K}_{n'}} \lambda_{K_m,\bar{K}_{m'}} \delta^2(K_n-\bar{K}_{n'}+K_m-\bar{K}_{m'}) c^\dagger_{K_n}c^\dagger_{K_m}c_{\bar{K}_{m'}}c_{\bar{K}_{n'}}\nonumber\\
    + &(K\leftrightarrow \bar{K})\nonumber\\
    = \sum_{nn'mm'} &V_{K_n-K_{n'}} \lambda_{K_n,K_{n'}} \lambda_{K_m,K_{m'}} \left(\delta_{nn'}\delta_{mm'}+\delta_{nm'}\delta_{n'm}-\delta_{nn'mm'}\right)
 c^\dagger_{K_n}c^\dagger_{K_m}c_{K_{m'}}c_{K_{n'}}\nonumber\\
    + &V_{K_n-K_{n'}} \lambda_{K_n,K_{n'}} \lambda_{\bar{K}_m,\bar{K}_{m'}} \left(\delta_{nn'}\delta_{mm'}+\delta_{nm}\delta_{n'm'}-\delta_{nn'mm'}\right) c^\dagger_{K_n}c^\dagger_{\bar{K}_m}c_{\bar{K}_{m'}}c_{K_{n'}}\nonumber\\
    - &V_{K_n-\bar{K}_{m'}} \lambda_{K_n,\bar{K}_{m'}} \lambda_{\bar{K}_m,K_{n'}} \left(\delta_{nn'}\delta_{mm'}+\delta_{nm}\delta_{n'm'}-\delta_{nn'mm'}\right) c^\dagger_{K_n}c^\dagger_{\bar{K}_m}c_{\bar{K}_{m'}}c_{K_{n'}} \nonumber\\
    + &(K\leftrightarrow \bar{K})
\end{align}
where $\delta_{nn'mm'}=1$ when $n=n'=m=m'$ and $\delta_{nn'mm'}=0$ otherwise. 
Now we transform to the angular momentum basis
\begin{align}
    c_{l,+} = \frac{1}{\sqrt{3}}\sum_{n=0,1,2} c_{K_n} \omega^{ln}\nonumber\\
    c_{l,-} = \frac{1}{\sqrt{3}}\sum_{n=0,1,2} c_{\bar{K}_n} \omega^{ln}
\end{align}
where $\omega=e^{i\frac{2\pi}{3}}$. The Hamiltonian becomes
\begin{align}
    H
    = \frac{1}{9}\sum_{ll'ss',nn'mm'} &V_{K_n-K_{n'}} \lambda_{K_n,K_{n'}} \lambda_{K_m,K_{m'}} \left(\delta_{nn'}\delta_{mm'}+\delta_{nm'}\delta_{n'm}-\delta_{nn'mm'}\right) \omega^{ln+sm-s'm'-l'n'}
 c^\dagger_{l+}c^\dagger_{s+}c_{s'+}c_{l'+}\nonumber\\ 
    + &V_{K_n-K_{n'}} \lambda_{K_n,K_{n'}} \lambda_{\bar{K}_m,\bar{K}_{m'}} \left(\delta_{nn'}\delta_{mm'}+\delta_{nm}\delta_{n'm'}-\delta_{nn'mm'}\right) \omega^{ln+sm-s'm'-l'n'} c^\dagger_{l+}c^\dagger_{s-}c_{s'-}c_{l'+}\nonumber\\ 
    - &V_{K_n-\bar{K}_{m'}} \lambda_{K_n,\bar{K}_{m'}} \lambda_{\bar{K}_m,K_{n'}}\left(\delta_{nn'}\delta_{mm'}+\delta_{nm}\delta_{n'm'}-\delta_{nn'mm'}\right) \omega^{ln+sm-s'm'-l'n'} c^\dagger_{l+}c^\dagger_{s-}c_{s'-}c_{l'+}\nonumber\\
    + &(K\leftrightarrow \bar{K})
    \label{fullH_L}
\end{align}
To get the effective 2-site and 3-state Potts model, in the following, we will focus on the terms that conserve the fermion number in each corner.
The first term of Eq.\ref{fullH_L} becomes
\begin{align}
    H_1 &=  \frac{1}{9} \sum_{ll'ss', nn'} V_{K_n-K_n'} |\lambda_{n'-n,+}|^2 \omega^{n(l-s')+n'(s-l')}c^\dagger_{l+}c^\dagger_{s+}c_{s'+}c_{l'+}\nonumber \\
    &+ \frac{1}{9} \sum_{ll'ss',nm} V(0)\omega^{n(l-l')+m(s-s')} c^\dagger_{l+}c^\dagger_{s+}c_{s'+}c_{l'+}
    - \frac{1}{9} \sum_{ll'ss',n} V(0) \omega^{n(l+s-s'-l')}c^\dagger_{l+}c^\dagger_{s+}c_{s'+}c_{l'+}\nonumber \\
    &= -\frac{1}{3}\sum_{ls,\Delta n,\alpha\beta} V_{\Delta n} \rho_{\alpha,+} \rho_{\beta,+} \omega^{(R_\alpha-R_\beta+l-l')\Delta n}\delta_3(l+s-s'-l') c^\dagger_{l+}c^\dagger_{s+}c_{s'+}c_{l'+} \nonumber\\
    &+\sum_{ls} V(0) c^\dagger_{l+}c^\dagger_{s+}c_{s+}c_{l+} + \frac{1}{3}\sum_{ll'ss'} V(0) \delta_3(l+s-s'-l') c^\dagger_{l+}c^\dagger_{s+}c_{s'+}c_{l'+} 
    %&= \left( \sum_{ls,\Delta n,\alpha\beta} V_{\Delta n} \rho_{\alpha,+} \rho_{\beta,+} \omega^{(R_\alpha-R_\beta + l - s)\Delta n}\right) \mathbb{1}_{K_M} \mathbb{1}_{\bar{K}_M}
\end{align}
where we have used
\be
    \lambda_{K_i,K_{i+n}} = \langle u(K_i) | u(K_{i+n}) \rangle = \sum_{\alpha} u^*_\alpha(K_i) u_\alpha(K_{i+n}) = \sum_\alpha |u_{\alpha}(K_0)|^2 \omega^{R_\alpha(n)} = \sum_\alpha \rho_{\alpha,+} \omega^{R_\alpha(n)}\equiv  \lambda_{n,+}
\ee
here $\alpha=0,1,2, ... ,9$ labels the 10 orbitals $(B_5, A_5, B_4, A_4, ..., B_1, A_1)$, while $R_\alpha=(0,1,1,2,2,3,3,4,4,5)$ is the in-plane displacement of the 10 orbitals from $B_5$ 
and
\be
    V_{n} = V(|\v K_0-\v K_n|) = V(G-G\delta_{n0}) =  V(G)+(V(0)-V(G))\delta_{n0}
\ee
As well, we have assumed that the mean-field ansatz to which this term is acted on has a conserved angular momentum (leading to $l = s$).
This term does not lift the degeneracy between different angular momentum channels.

The second term of Eq.\ref{fullH_L} is
\begin{align}
    H_2 =& 
    +\frac{1}{9}\sum_{ll'ss',n\Delta n} V_{\Delta n} \lambda_{\Delta n,+} \lambda_{\Delta n,-} \omega^{(l+s)\Delta n} \omega^{n(l+s-s'-l')}c^\dagger_{l+}c^\dagger_{s-}c_{s'-}c_{l'+}\nonumber\\
    & +\frac{1}{9}\sum_{ll'ss', nm} V(0) \omega^{n(l-l')+m(s-s')}c^\dagger_{l+}c^\dagger_{s-}c_{s'-}c_{l'+} - \frac{1}{9}\sum_{ll'ss',n} V(0)\omega^{n(l+s-s'-l')}c^\dagger_{l+}c^\dagger_{s-}c_{s'-}c_{l'+}\nonumber\\
    =& +\frac{1}{3}\sum_{ll'ss',\Delta n,\alpha\beta} V(G) \rho_{\alpha,+} \rho_{\beta,-} \omega^{(R_\alpha+R_\beta+l+s)\Delta n} \delta_3(l+s-s'-l')c^\dagger_{l+}c^\dagger_{s-}c_{s'-}c_{l'+}\nonumber\\
    & + \frac{1}{3}\sum_{ls,\alpha\beta} (V(0)-V(G))\delta_3(l+s-s'-l')c^\dagger_{l+}c^\dagger_{s-}c_{s'-}c_{l'+}\nonumber\\
    & + \sum_{ls} V(0) c^\dagger_{l+}c^\dagger_{s-}c_{s-}c_{l+} - \frac{1}{3}\sum_{ll'ss'} V(0)\delta_3(l+s-s'-l')c^\dagger_{l+}c^\dagger_{s-}c_{s'-}c_{l'+}
\end{align}
after a mean-field decoupling, only the first line is non-trivial
\begin{align}
    H_2&=
    \sum_{ls,\alpha\beta} V(G) \rho_{\alpha,+} \rho_{\beta,-} \delta_3(R_\alpha+R_\beta+l+s) \langle c^\dagger_{l+}c_{l+} \rangle \langle c^\dagger_{s-}c_{s-}\rangle + const.\nonumber\\
    &=\frac{2}{3} V(G) \sum_{\alpha\beta} \rho_{\alpha,+} \rho_{\beta,-} \cos\left(\theta_1+\theta_2+\frac{2(R_\alpha+R_\beta)\pi}{3}\right) + const.
\end{align}
This is the Hatree term for inter-MBZ-corner interaction. When the orbital weight $\rho_\alpha$ is polarized to $B_5$ $(\alpha=0)$, this term is maximized for $l+s\equiv-(\alpha+\beta)\equiv 0$ (mod 3). Note that $l+s$ (mod 3) is associated with the Chern number. So this term penalizes $C=0$. As soon as the orbital weights shift to $B_4(\alpha=\beta=1)$, this term penalizes $s+l\equiv1$(mod 3) instead. 

The third line of Eq.~\ref{fullH_L} becomes
\begin{align}
    H_3=& -\frac{1}{9}\sum_{ll'ss',nm} V_{K_n-\bar{K}_m} |\lambda_{K_n,\bar{K}_m}|^2 \omega^{n(l-l')+m(s-s')} c^\dagger_{l+}c^\dagger_{s-}c_{s'-}c_{l'+}\nonumber \\
    & -\frac{1}{9}\sum_{ll'ss',nn'} V_{K_n-\bar{K}_{n'}} \lambda_{K_n,\bar{K}_{n'}} 
    \lambda_{\bar{K}_n,K_{n'}}\omega^{n(l+s)-n'(l'+s')} c^\dagger_{l+}c^\dagger_{s-}c_{s'-}c_{l'+}\nonumber\\
    & +\frac{1}{9}\sum_{ll'ss',n} V_{K_n-\bar{K}_{n}} |\lambda_{K_n,\bar{K}_{n}}|^2 \omega^{n(l+s-s'-l')} c^\dagger_{l+}c^\dagger_{s-}c_{s'-}c_{l'+}\nonumber\\
    =&-\frac{1}{3}\sum_{ll'ss',\Delta n} V'_{\Delta n} |\lambda^\prime_{\Delta n}|^2 \omega^{(s-s')\Delta n} \delta_3(l+s-s'-l') c^\dagger_{l+}c^\dagger_{s-}c_{s'-}c_{l'+}\nonumber\\
    &-\frac{1}{3}\sum_{ll'ss',\Delta n} V'_{\Delta n} \lambda_{\Delta n}^{\prime2} \omega^{(l'+s')\Delta n} \delta_3(l+s-s'-l') c^\dagger_{l+}c^\dagger_{s-}c_{s'-}c_{l'+}\nonumber\\
    &+\frac{1}{3}\sum_{ll'ss'} V'_0 |\lambda^\prime_{0}|^2 \delta_3(l+s-s'-l') c^\dagger_{l+}c^\dagger_{s-}c_{s'-}c_{l'+}\nonumber\\
    =&-\sum_{ll'ss'} V(K)  |\lambda^\prime_1|^2\delta(s-s') \delta_3(l+s-s'-l') c^\dagger_{l+}c^\dagger_{s-}c_{s'-}c_{l'+}\nonumber\\
    &-\frac{1}{3}\sum_{ll'ss'} V(K)   \sum_{\Delta n}\lambda_{\Delta n}^{\prime2}\omega^{(l'+s')\Delta n} \delta_3(l+s-s'-l') c^\dagger_{l+}c^\dagger_{s-}c_{s'-}c_{l'+}\nonumber\\
    &-\frac{1}{3}\sum_{ll'ss'} (2V(2K)|\lambda^\prime_0|^2-V(K)|\lambda^\prime_1|^2-V(K)\lambda_0^{\prime2})  \delta_3(l+s-s'-l') c^\dagger_{l+}c^\dagger_{s-}c_{s'-}c_{l'+}\nonumber\\
    &+\frac{1}{3}\sum_{ll'ss'} V(2K) |\lambda^\prime_{0}|^2 \delta_3(l+s-s'-l') c^\dagger_{l+}c^\dagger_{s-}c_{s'-}c_{l'+}
    \label{term3}
\end{align}
where
\be
    V'_{n} = V(K_i-\bar{K}_{n+i})= V(K_i+K_{n+i}) = V(K)+(V(2K)-V(K))\delta_{n0}
\ee
\be
     \lambda_{K_i,\bar{K}_{i+n}} = \langle u(K_i) | u(\bar{K}_{i+n}) \rangle = \sum_{\alpha} u^*_\alpha(K_i) u_\alpha(\bar{K}_{i+n}) = \sum_\alpha u^*_\alpha(K_0)u_\alpha(\bar{K}_0) \omega^{R_\alpha n} = \sum_\alpha \rho'_{\alpha} (-1)^{R_\alpha} \omega^{R_\alpha n}\equiv \lambda'_n
\ee
At the mean-field level, only the second line of Eq.~\ref{term3} is non-trivial, which reduces to
\begin{align}
    H_3&= -\frac{1}{3} V(K) \sum_{ls, \Delta n}\lambda^{\prime2}_{\Delta n} \omega^{(l+s)\Delta n} \langle c^\dagger_{l+}c_{l+} \rangle \langle c^\dagger_{s-}c_{s-}\rangle + const.\nonumber\\
    &= -\frac{1}{3} V(K) \sum_{ls,\alpha\beta} \rho_{\alpha}^{\prime} \rho^{\prime}_{\beta} (-1)^{R_\alpha+R_\beta} \sum_{\Delta n}\omega^{(s+l+R_\alpha+R_\beta)\Delta n} \langle c^\dagger_{l+}c_{l+} \rangle \langle c^\dagger_{s-}c_{s-}\rangle\nonumber\\
    &= -V(K) \sum_{ls,\alpha\beta} \rho_{\alpha}^{\prime} \rho^{\prime}_{\beta} (-1)^{R_\alpha+R_\beta} \delta_3(s+l+R_\alpha+R_\beta) \langle c^\dagger_{l+}c_{l+} \rangle \langle c^\dagger_{s-}c_{s-}\rangle\nonumber\\
    &= -\frac{2}{3} V(K) \sum_{\alpha\beta} \rho_{\alpha}^{\prime} \rho^{\prime}_{\beta} (-1)^{R_\alpha+R_\beta} \cos\left(\theta_1+\theta_2 + \frac{2\pi (R_\alpha+R_\beta)}{3}\right) + const.
\end{align}
This is the Fock term between $K_M$ and $\bar{K}_M$.

The total Hamiltonian 
\begin{align}
    H = \frac{2}{3} \sum_{\alpha\beta}  \left(V(G) \rho_{\alpha} \rho_{\beta} -V(K) \rho_{\alpha}^{\prime} \rho^{\prime}_{\beta} (-1)^{R_\alpha+R_\beta} \right) \cos\left(\theta_1+\theta_2 + \frac{2\pi (R_\alpha+R_\beta)}{3}\right) + const.
    \label{Htotal}
\end{align}
At a large displacement field and small twist angles, although there is significant hybridization between $B_5$ and $B_4$ at BZ corners, the density is still predominantly on $B_5$ orbital. As a simple consideration, consider the extreme scenario where the charge is polarized to the $B_5$ orbital, $\rho_0=\rho'_0=1$, then
\begin{align}
    H_2 =\frac{2}{3} \left(V(G)-V(K)\right)  \cos\left(\theta_1+\theta_2\right) + const.
\end{align}
Since $V(K)>V(G)$, Fock term dominates. Then $C\equiv l+s=\frac{3(\theta_1+\theta_2)}{2\pi}\equiv 0$ (mod 3) is lower in energy than $C=1$ and $C=-1$. We note that this finding is inconsistent with Berry curvature rounding. Since for layer-polarized Bloch function, the Berry flux in MBZ is zero. This issue will be resolved in the Appendix~\ref{app:SCring}.
%This is consistent with the Berry curvature rounding argument.

\begin{figure}[h]
    \centering
    \includegraphics[width=0.7\linewidth]{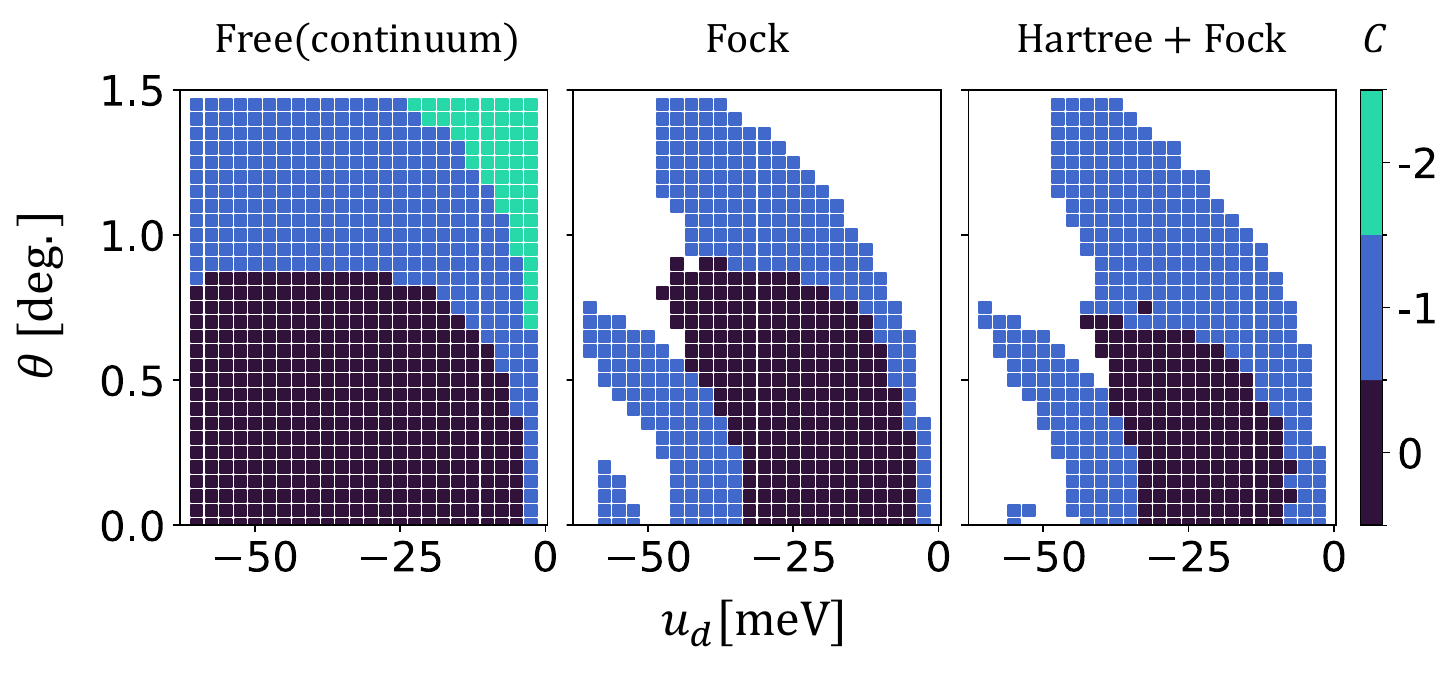}
    \includegraphics[width=0.7\linewidth]{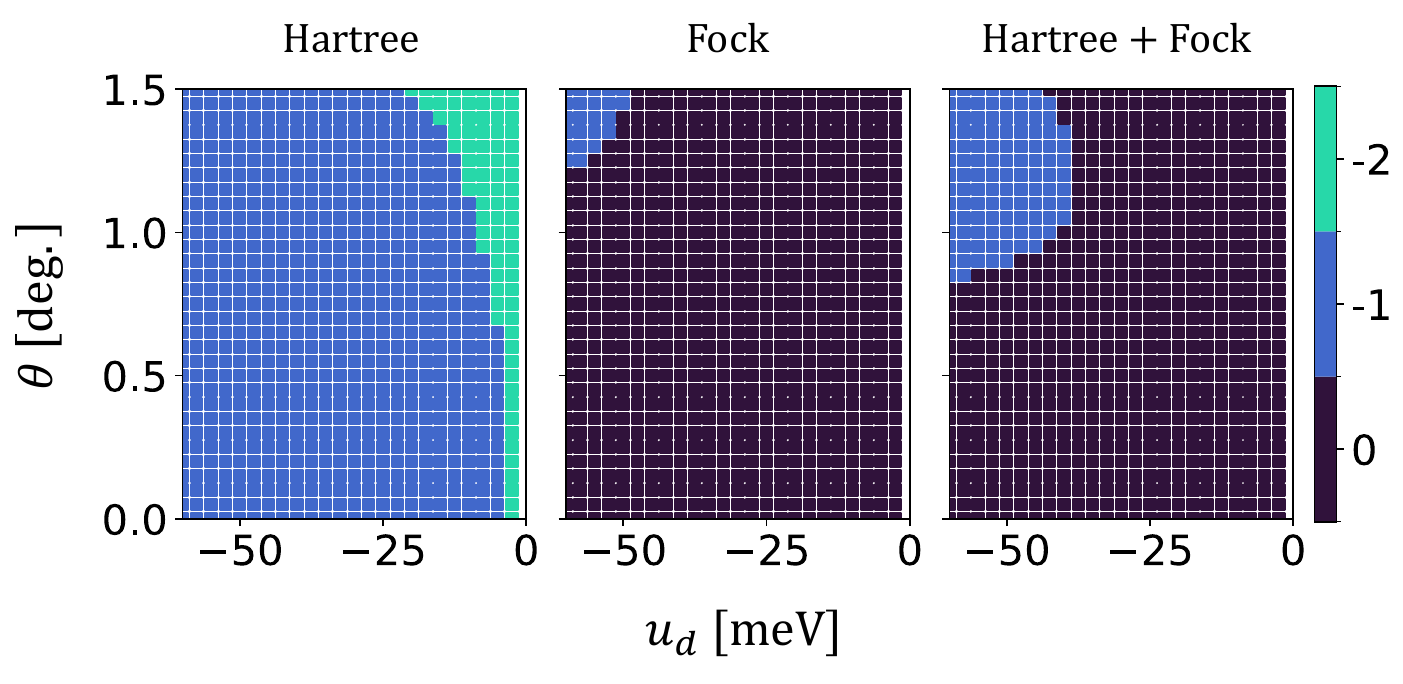}
    \caption{Comparison between microscopic mean field and simplified model based on MBZ corners. Top row: phase diagram from microscopic mean-field. (Same as Fig.~\ref{hf_pd_compare} in main text.) Bottom row: Hartree, Fock and Hartree+Fock phase diagrams based on the simplified pseudopotential model Eq.~\ref{pseudo_final}.
    }
   \label{pd_corner}
\end{figure}

At slightly larger twist angles, density starts to leak into the $B_4$ orbital. The leading correction comes from the $\rho_{0}\rho_{1}$ term. 
We take the assumption of approximate $C_6$ symmetry, which means
\begin{equation}
    \rho^\prime_a = |u^*_a(K_0)u_a(\bar{K}_0)| = \rho_a
\end{equation}
As a result, the Fock term gets suppressed by the inter-layer terms with $a\neq b$ (mod 2), which may allow the Hartree term to take over and favor $C=\pm1$. 
This corresponds to the intermediate twist angle regime ($0.6^\circ <\theta <0.9^\circ$), as seen in Fig. \ref{pd_corner}.

%\zhihuan{Need to generate a phase diagram based on Eq.~\ref{Htotal}.}

\begin{figure}[h]
    \centering
    \includegraphics[width=0.5\linewidth]{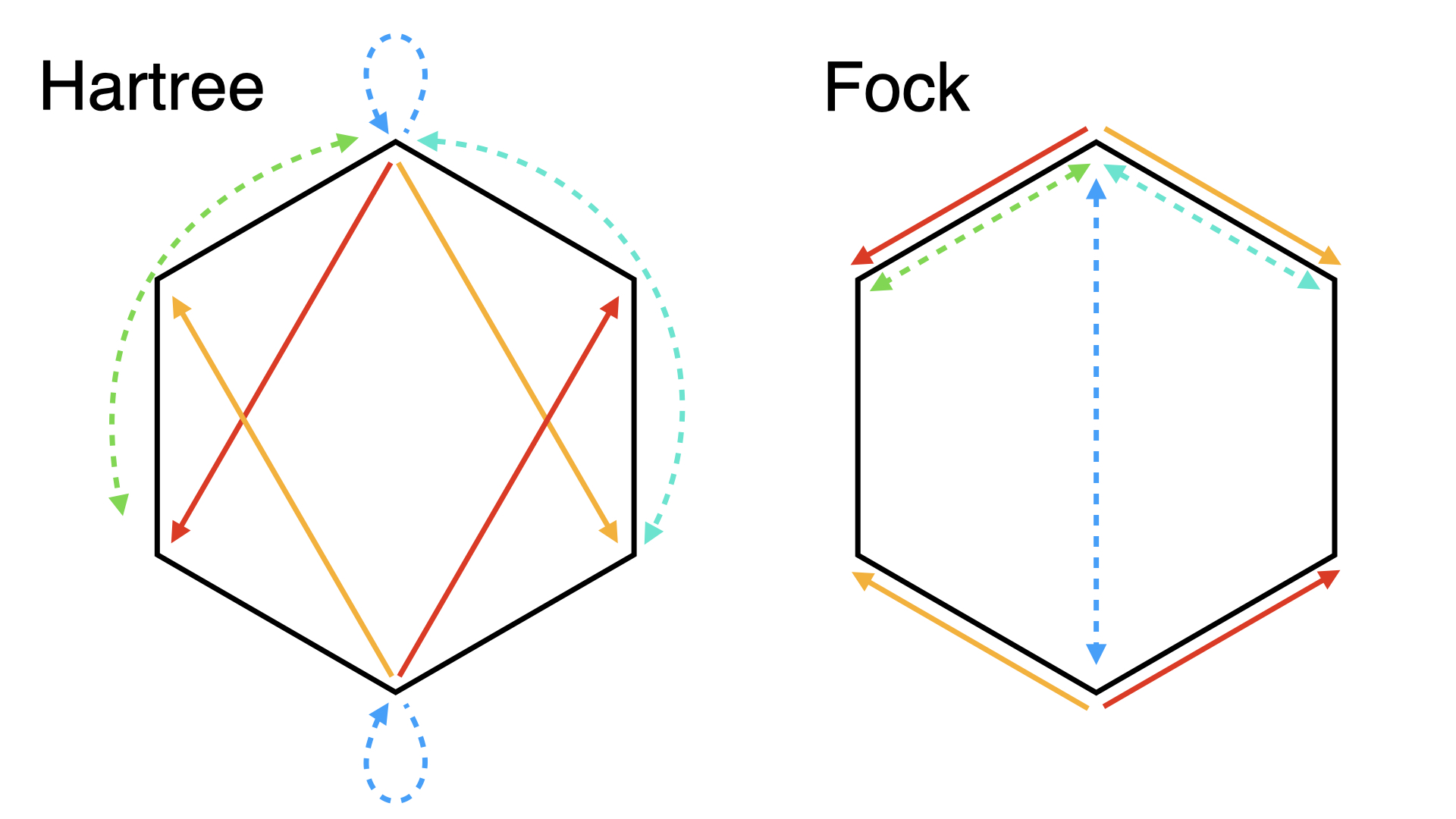}
    \caption{Hartree and Fock terms in the pseudopotential model. The red and orange arrows are non-trivial inter-corner scattering processes. The dotted arrows are scattering processes that do not lift degeneracy between different angular momentum channels.} %\adarsh{placeholder for now; can modify as we need it} \zhihuan{the coloring is inconsistent. Let's color the same process with the same color.}}
   \label{fig_hf_hex}
\end{figure}

The important terms that determine the Chern number can be seen more explicitly, by considering the responsible terms with the form factors intact.
From Fig.~\ref{fig_hf_hex} the Hartree term is
\begin{align}
    H_2 
    =& V_{K_n-K_{n'}} \lambda_{K_n,K_{n'}} \lambda_{\bar{K}_m,\bar{K}_{m'}} \delta_{nm}\delta_{n'm'} \omega^{ln+sm-s'm'-l'n'} c^\dagger_{l+}c^\dagger_{s-}c_{s'-}c_{l'+}\nonumber\\ 
    =& 
    \frac{1}{9}\sum_{ll'ss'}\sum_{n,\Delta n=\pm1} V_{\Delta n} \lambda_{\Delta n,+} \lambda_{\Delta n,-} \omega^{(l+s)\Delta n} \omega^{n(l+s-s'-l')}c^\dagger_{l+}c^\dagger_{s-}c_{s'-}c_{l'+} + const.\nonumber\\
    \sim & 
    \frac{1}{3}\sum_{ll'ss'} V(G) \left(\lambda_{0,1} \lambda_{\bar{0},\bar{1}} \omega^{(l+s)} + h.c. \right) \delta_3(l+s-s'-l')c^\dagger_{l+}c^\dagger_{s-}c_{s'-}c_{l'+}\nonumber\\
    =& 
    \frac{1}{3}\sum_{ls} V(G) 2 \Re\left(\lambda_{0,1}^2 \omega^{(l+s)}\right) \langle c^\dagger_{l+}c_{l+}\rangle \langle c^\dagger_{s-}c_{s-}\rangle,
    \label{compact}
\end{align}
and the Fock term is
\begin{align}
    H_3
    =&
    -V_{K_n-\bar{K}_{m'}} \lambda_{K_n,\bar{K}_{m'}} \lambda_{\bar{K}_m,K_{n'}}\delta_{nm}\delta_{n'm'} \omega^{ln+sm-s'm'-l'n'} c^\dagger_{l+}c^\dagger_{s-}c_{s'-}c_{l'+}\nonumber\\
    =&-\frac{1}{9}\sum_{ll'ss'}\sum_{n,\Delta n=\pm1} V'_{\Delta n} \lambda_{\Delta n}^{\prime2} \omega^{(l'+s')\Delta n} \omega^{n(l+s-s'-l')} c^\dagger_{l+}c^\dagger_{s-}c_{s'-}c_{l'+}+const.\nonumber\\
    \sim& 
    -\frac{1}{3}\sum_{ll'ss'} V(K) \left(\lambda_{0,\bar{1}}\lambda_{\bar{0},1} \omega^{(l'+s')} + h.c. \right) \delta_3(l+s-s'-l') c^\dagger_{l+}c^\dagger_{s-}c_{s'-}c_{l'+}\nonumber\\
    =& 
    -\frac{1}{3}\sum_{ls} V(K) 2 \Re\left(\lambda_{0,\bar{1}}^2 \omega^{(l+s)}\right) \langle c^\dagger_{l+}c_{l+}\rangle \langle c^\dagger_{s-}c_{s-}\rangle
\end{align}
Thus, the total Hamiltonian becomes
\begin{align}
    H_{MF} = H_2+H_3 = -\frac{2}{3} \Re\left[\left(-V(G)\lambda_{0,1}^2+ V(K)\lambda_{0,\bar{1}}^2\right) e^{i(\theta_1+\theta_2)}\right].
    \label{pseudo_final}
\end{align}
The ground state of Potts model has a corresponding Chern number
\begin{align}
    C \equiv \frac{3(\theta_1+\theta_2)}{2\pi} \equiv -\frac{3 \arg\left(-V(G)\lambda_{0,1}^2+ V(K)\lambda_{0,\bar{1}}^2\right)}{2\pi} \text{ (mod 3)}.
\end{align}

%\adarsh{missing a factor of $2/3$ on the right hand side of D22?}

We present in Fig.~\ref{pd_corner} the Hartree, Fock, and Hartree+Fock phase diagrams of this simplified pseudopotential model.
As seen, the Hartree solution favors $C = -1$ over a wide range of parameter space, while the Fock term, prefers $C = 0$.
%Nonetheless, in the approximate range of the experimentally relevant twist angles and displacement field, an island of $C = -1$ appears in this simplified pseudopotential model, broadly consistent with our complete Hartree-Fock studies in the main text.

\section{Failure of the pseudopotential treatment}
In the previous section, we only consider the degree of freedom at the MBZ corners $K$ and $\bar{K}$. This treatment is justified by asserting that the wavefunction evolves smoothly so that in a small region close to each corner of MBZ, the wavefunction of the normalized band is roughly the same as that on the exact corner. The extension of this region is defined by the relative strength of interaction and kinetic energy, namely the radius $q\sim U/v(K)$, where U is the Coulomb interaction and $v(K)$ is the velocity at the corner. Within this region, the Coulomb energy dominates over kinetic energy, so that the Bloch function is approximately an equal-weight superposition of three continuum band wavefunctions $c_{\v K_0+\v k}, c_{\v K_1+\v k}, c_{\v K_2+\v k}$.

%However, this stops being a legitimate approximation when the wavefunction varies violently within the same patch. In this section, we describe the leading corrections in this scenario and make connection to the Berry curvature rounding argument in Appendix~\ref{FockBerry}.

%In Appendix.~\ref{app_alternative_gl_model}, we dropped the intra-patch Fock term. Naively this term does not lift the degeneracy between three Chern numbers. But this is only correct under the assumption that the wavefunction in each patch is exactly uniform.

The Chern number is eventually determined by symmetry index on the exact $\v K$ and $\bar{\v K}$ points. So we focus on the mean-field Hamiltonian for these momenta
\begin{align}
    H_{Fock}(\v K_0)
    &= -\sum_{\v k',\v G} V(\v K_0+\v G-\v k') \lambda_{\v k', \v K_0+\v G} \lambda_{\v K_0, \v k'-\v G} \langle c^\dagger_{\v k'}c_{\v k'-\v G} \rangle  c^\dagger_{\v K_0} c_{\v K_0+\v G}
\end{align}
To show the correction to the pseudopotential model, we focus on the scattering matrix element between $\v K_0$ and $\v K_1$. This corresponds to the process marked by orange arrows in Fig.~\ref{fig_hf_hex}(b). Therefore we restrict to $\v k'= \bar{\v K}_0+\v q'$, $\v K_0+\v G=\v K_1$
\begin{align}
    H_{Fock}(\v K_0, \v K_1)
    &= -\sum_{\v q'} V(\v K_1-\bar{\v K}_0-\v q') \lambda_{\bar{\v K}_0+\v q', \v K_1} \lambda_{\v K_0, \bar{\v K}_1+\v q'} \langle c^\dagger_{\bar{\v K}_0+\v q'}c_{\bar{\v K}_1+\v q'} \rangle  %c^\dagger_{\v K_0} c_{\v K_1}
\end{align}
Now rewrite this using the basis of the Fock-renormalized band
\be
   \psi^n_{\v k} = \sum_{\v G} v^{n*}_{\v G}(\v k) c_{\v k+\v G} \equiv \sum_{\v G} v^{n*}(\v k+\v G) c_{\v k+\v G}
\ee
so that $\psi^n_{\v k+\v g}=\psi^n_{\v k}$. Here $n$ is the band index, which will be suppressed since we are focusing on the mean-field ansatz of an insulator, where only the lowest band is filled. Then using the inverse transform,
\be
   c_{\v k+\v G} = \sum_{n} v^{n}_{\v G}(\v k) \psi^n_{\v k} = \sum_{n} v^{n}(\v k+\v G) \psi^n_{\v k}
\ee
The Fock Hamiltonian (associated with these two corners) becomes
\begin{align}
    \label{E1}
    H_{Fock}(\v K_0, \v K_1)
    &= -\sum_{\v q'} V(\v K_1-\bar{\v K}_0-\v q') \lambda_{\bar{\v K}_0+\v q', \v K_1} \lambda_{\v K_0, \bar{\v K}_1+\v q'}  v^*(\bar{\v K}_0+\v q') v(\bar{\v K}_1+\v q') \langle \psi^\dagger_{\bar{\v K}_0+\v q'}\psi_{\bar{\v K}_1+\v q'} \rangle\nonumber\\
    &= -\sum_{\v q'} V(\v K_1-\bar{\v K}_0-\v q') \lambda_{\bar{\v K}_0+\v q', \v K_1} \lambda_{\v K_0, 
    \bar{\v K}_1+\v q'}  v^*(\bar{\v K}_0+\v q') v(\bar{\v K}_1+\v q')\nonumber\\
    &= -\sum_{\v q'} V(\v K_1-\bar{\v K}_0-\v q') 
    \sum_{\alpha\beta} v^*(\bar{\v K}_0+\v q') u^*_\alpha(\bar{\v K}_0+\v q')u_\alpha(\v K_1) u^*_\beta(\v K_0)u_\beta(\bar{\v K}_1+\v q')  v(\bar{\v K}_1+\v q')\nonumber\\
    &= -\sum_{\v q'} V(\v K_1-\bar{\v K}_0-\v q') 
    \sum_{\alpha\beta} \xi_\alpha^*(\bar{\v K}_0+\v q')u_\alpha(\v K_1) u^*_\beta(\v K_0) \xi_\beta(\bar{\v K}_1+\v q')\nonumber\\
    &\equiv -\sum_{\v q'} V(\v K_1-\bar{\v K}_0-\v q') \sum_{\alpha\beta}F_{\alpha\beta}(\v q)u_\alpha(\v K_1) u^*_\beta(\v K_0)
\end{align}
where we have used the Bloch function of the renormalized band
\be
    \xi_{\alpha,\v G}(\v k) = v_{\v G}(\v k) u_\alpha(\v k+\v G) \text{ for } \v k \in BZ
\ee
which has been extended to $\v k$ beyond BZ by defining $\xi_{\alpha,\v G}(\v k) \equiv \xi_{\alpha}(\v k+\v G)$
\be
    \xi_{\alpha}(\v k) = v(\v k) u_\alpha(\v k) \text{ for } \forall \v k
\ee
Expanding $F(\v q')$ around $\v q'=0$,
\begin{align}
    \label{E2}
    F_{\alpha\beta}(\v q')
    &= \xi_\alpha^*(\bar{\v K}_0+\v q') \xi_\beta(\bar{\v K}_1+\v q')\nonumber\\
    &\approx \langle \xi(\bar{\v K}_0+\v q')|\xi(\bar{\v K}_0)\rangle \langle \xi(\bar{\v K}_1)|\xi(\bar{\v K}_1+\v q')\rangle \xi_\alpha^*(\bar{\v K}_0) \xi_\beta(\bar{\v K}_1)\nonumber\\
    &= \tilde{\lambda}_{\bar{\v K}_0+\v q',\bar{\v K}_0} \tilde{\lambda}_{\bar{\v K}_1,\bar{\v K}_1+\v q'} \xi_\alpha^*(\bar{\v K}_0) \xi_\beta(\bar{\v K}_1)\nonumber\\
    &= |\tilde{\lambda}_{\bar{\v K}_0+\v q',\bar{\v K}_0}|^2 \xi_\alpha^*(\bar{\v K}_0) \xi_\beta(\bar{\v K}_1)\nonumber\\
    &\approx e^{-\tilde{g}_{\mu\nu} q'_\mu q'_\nu} \xi_\alpha^*(\bar{\v K}_0) \xi_\beta(\bar{\v K}_1)
\end{align}
where $\tilde{\lambda}$ and $\tilde{g}$ are the form factor and quantum metric of the normalized band. In the second line, we made the approximation by only keeping the projection of the two vectors $\xi_\alpha^*(\bar{\v K}_0+\v q')$ and $\xi_\beta(\bar{\v K}_1+\v q')$ to their counterpart at $\v q'=0$. The result should be interpreted as the ``fidelity" of representing a corner region with Bloch function exactly on the high symmetry points. The exponential suppression shows that this treatment is crude for the ansatz that violates Berry curvature rounding.

\section{$k$-space superconducting ring model and connection to Berry curvature rounding}
\label{app:SCring}
Although we aim to eventually derive a simplified model defined on the BZ corners, the microscopic low energy degrees of freedom are not restricted to the corners. Instead, they live on the entire BZ boundary. We demonstrate in this section, that the emergent theory for these fluctuations is essentially an XY field coupled to a background gauge field, analogous to a superconducting ring (along BZ boundary) under a background magnetic field (parent Berry curvature). Then, the rounding of Berry curvature is analogous to the Little-Parks effect, where the winding of XY order parameter around a narrow superconducting ring is given by rounding the magnetic flux piercing the ring. This analogy becomes exact when the long-wavelength component of interaction dominates.

Consider the mean-field phase diagram as the interaction is turned up. At weak interaction, we get a fermi liquid with no translation symmetry breaking. With moderate interaction, the translation breaks weakly, and we find a semi-metal with compensating particle and hole pockets. The mean-field periodic potential is not strong enough to open a global(indirect) band gap. Nevertheless, the Chern number for the lowest mean field band is still well-defined in this regime. If we keep increasing the interaction, no gap closing happens until the band gap fully opens. Unless there is a discontinuous transition at the mean-field level, the resulting insulating phase at strong interaction must have the same Chern number as in the semi-metal phase at moderate interaction. Therefore, one can predict the fate of Chern number at strong interaction regime by studying the intermediate regime, where interaction is just enough to open a direct band gap. 

\begin{figure}[h]
    \centering
    \includegraphics[width=0.7\linewidth]{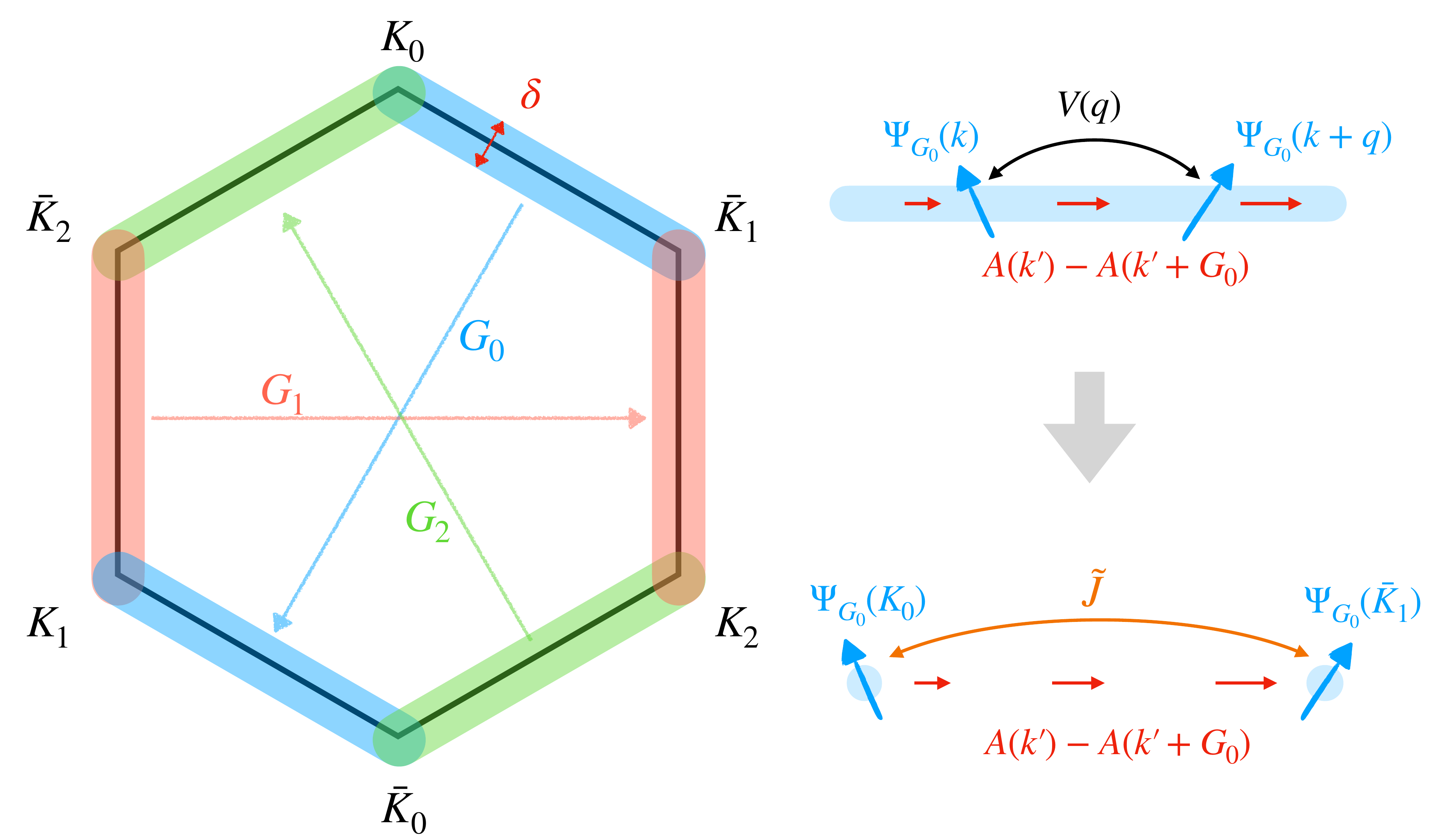}
    \caption{Superconducting ring model. The interaction between corners is generated by integrating out the superconducting wire connecting them.}
   \label{SCring}
\end{figure}

In this regime, $V_G\ll K v_F$, the active degree of freedom is not just on the corners of BZ, but instead on a hexagonal ring along the BZ boundary, with a width of $\delta\sim V_G/v_F\ll K$. That is to say, in the language of the pseudopotential model, the $K$ and $\bar{K}$ corner regions are highly anisotropic.
The structure of the active hexagon ring consists of two types of regions: (1) corners, where 3 momenta are hybridized, and (2) edges, where only 2 momenta are hybridized significantly. Note that region (1) has a radius of $\delta$, which limits its area to be $O(\delta^2)$, much smaller than region (2), whose area is of $O(K\delta)$. Therefore we will minimize the energy in region (2) first.

For a hexagonal BZ, there are three segments in the region (2). We will focus on one of them, the edge between $K_0$ and $\bar{K}_1$, for example.
The Fock term is
\begin{align}
    H_{Fock}
    &= -\sum_{\v k,\v q} V(\v q) \lambda_{\v k, \v k + \v q} \lambda^*_{\v k+\v G, \v k + \v q +\v G} \langle c^\dagger_{\v k + \v q +\v G}c_{\v k +\v q} \rangle  \langle c^\dagger_{\v k} c_{\v k+\v G} \rangle\nonumber \\
    &= -\sum_{\v k,\v q} V(\v q) \lambda_{\v k, \v k + \v q} \lambda^*_{\v k+\v G, \v k + \v q +\v G} \Psi^*_{\v G}(\v k +\v q) \Psi_{\v G}(\v k)
\end{align}
Here for unscreened Coulomb $V(\v q)\sim 1/q$, and $\v G=\v K_1-\v K_0$. We note this theory is a 1d superconducting wire with Josephson coupling in k-space. Now expand the form factor for small $q$
\begin{equation}
    \lambda_{\v k,\v k+\v q}=\langle \v k| \v k+\v q\rangle \approx \exp\left\{- \frac{1}{2}g_{\mu\nu}(\v k)q_{\mu}q_{\nu} \right\} \exp\left\{i \int_{\v k}^{\v k+\v q} d \v q' \cdot A(\v q')\right\}
    \label{formfactor0}
\end{equation}
Then the small-$q$ contribution from Fock term becomes
\begin{align}
    H_{Fock}
    &= -\sum_{\v k} \sum_{\v q\ll K} V(\v q) \exp\left\{- \frac{1}{2}\left(g_{\mu\nu}(\v k)+g_{\mu\nu}(\v k+\v G)\right) q_{\mu}q_{\nu} \right\} \exp\left\{i\int_{\v k}^{\v k+\v q} d\v q' \cdot (A(\v q') -A(\v q'+\v G)) \right\} \Psi^*_{\v G}(\v k +\v q) \Psi_{\v G}(\v k)\nonumber \\
    &= -\sum_{\v k} \sum_{\v q\ll K} \tilde{V}(\v q, \v k) \exp\left\{i\int_{\v k}^{\v k+\v q} d\v q' \cdot (A(\v q') -A(\v q'+\v G)) \right\} \Psi^*_{\v G}(\v k +\v q) \Psi_{\v G}(\v k)
    \label{SCH1}
\end{align}
where we define $\tilde{V}(\v q, \v k) =  V(\v q) \exp\left\{- \frac{1}{2}\left(g_{\mu\nu}(\v k)+g_{\mu\nu}(\v k+\v G)\right) q_{\mu}q_{\nu} \right\}  $. This
 is essentially a 1d Josephson array under a background gauge field. Note the Josephson coupling $\tilde{V}(\v q, \v k)$ decays rapidly with $\v q$ as a Gaussian. Therefore, this problem is local in momentum space. This justifies the small-$q$ expansion in Eq.~\ref{formfactor0}. In other words, the small-$q$ coupling in Eq.~\ref{SCH1} indeed dominates the Fock term and should be minimized first. Then we find the ground state phase configuration
\begin{equation}
    \Psi_{\v G}(\v k) = \Psi_{\v G}(\v K_0) \exp\left\{i\int_{\v K_0}^{\v k} d\v q' \cdot \left(A(\v q') -A(\v q'+\v G)\right)\right\}
\end{equation}
Plugging in $\v k=\bar{\v K}_1$, we find a constraint between $\Psi_{\v G}(\v K_0)\equiv \Psi(\v K_0, \v K_0+\v G)=\Psi(\v K_0, \v K_1)$ and $\Psi_{\v G}(\bar{\v K}_1)=\Psi(\bar{\v K}_1, \bar{\v K}_1+\v G)\equiv \Psi(\bar{\v K}_1, \bar{\v K}_0)=\Psi(\bar{\v K}_0, \bar{\v K}_1)^*$. Namely,
\begin{align}
    \Psi(\bar{\v K}_0, \bar{\v K}_1)^* &= \Psi(\v K_0, \v K_1) \exp\left\{i\int_{\v K_0}^{\bar{\v K}_1} d\v q' \cdot \left(A(\v q') -A(\v q'+\v G)\right)\right\}\nonumber \\
    &= \Psi(\v K_0, \v K_1) \exp\left\{i\int_{\v K_0}^{\bar{\v K}_1} + \int_{\bar{\v K}_0}^{\v K_1} d\v q' \cdot A(\v q')\right\}
\end{align}
So the sum of phases of $\Psi(\bar{\v K}_0, \bar{\v K}_1)$ and $\Psi(\v K_0, \v K_1)$ is fixed.
\begin{align}
    \operatorname{Im} \left(\ln \Psi(\bar{\v K}_0, \bar{\v K}_1) + \ln(\Psi(\v K_0, \v K_1)\right) = -\left\{\int_{\v K_0}^{\bar{\v K}_1} + \int_{\bar{\v K}_0}^{\v K_1}\right\} d\v q' \cdot A(\v q')
\end{align}
It is easy to show that the gauge-invariant $C_3$ indices at high symmetry points are given by the sum of the three phases, which is also gauge-invariant
\begin{align}
    C_3(\v K) = \frac{1}{2\pi} \operatorname{Im} \left(\ln \Psi(\v K_0, \v K_1)+\ln \Psi(\v K_1, \v K_2)+\ln \Psi(\v K_2, \v K_0)\right)\nonumber \\
    C_3(\bar{\v K}) = \frac{1}{2\pi} \operatorname{Im} \left(\ln \Psi(\bar{\v K}_0, \bar{\v K}_1)+\ln \Psi(\bar{\v K}_1, \bar{\v K}_2)+\ln \Psi(\bar{\v K}_2, \bar{\v K}_0)\right)
\end{align}
Then the Chern number will be
\begin{align}
    C &= C_3(\v K)+C_3(\bar{\v K})\\
    &= \operatorname{Im} \left(\ln \Psi(\v K_0, \v K_1)+\ln \Psi(\v K_1, \v K_2)+\ln \Psi(\v K_2, \v K_0) + \ln \Psi(\bar{\v K}_0, \bar{\v K}_1)+\ln \Psi(\bar{\v K}_1, \bar{\v K}_2)+\ln \Psi(\bar{\v K}_2, \bar{\v K}_0)\right)\nonumber \\
    & = -\frac{1}{2\pi} \left\{\int_{\v K_0}^{\bar{\v K}_1} + \int_{\bar{\v K}_1}^{\v K_2} + \int_{\v K_2}^{\bar{\v K}_0} + \int_{\bar{\v K}_0}^{\v K_1}+ \int_{\v K_1}^{\bar{\v K}_2}+ \int_{\bar{\v K}_2}^{\v K_0}\right\} d\v q' \cdot A(\v q') = \frac{\Phi_{BZ}}{2\pi}
    \label{Chernnumber}
\end{align}
To obtain this result, we do not even need to assume $C_2$. Obviously, this result makes sense only when $\Phi_{BZ}/2\pi \in Z$ since the Chern number must be an integer. This is because we have assumed the SC wire to be at its ground state. However, when the Berry flux is not an integer multiple of $2\pi$, there is frustration. The correct treatment is integrating out the phase fluctuation in the middle of the wire and constructing a theory for the order parameter at $\v K$ and $\bar{\v K}$ (see Fig.~\ref{SCring}(b)).

We may start by writing down the effective coupling Hamiltonian for the two endpoints of a 1D SC wire. This is done by considering the energy cost of twisting the phases on two ends $\theta_{1,2}$. The leading gauge-invariant term is
\begin{align}
    H_{Fock}[\Psi_1, \Psi_2] &= -J|\Psi|^2 \cos\left(\theta_1-\theta_2+\int_{1}^{2} d\v q' \cdot (A(\v q') -A(\v q'+\v G))\right)\nonumber \\
    &= -\frac{J}{2} \Psi_1\Psi^*_2\exp\left\{i\int_{1}^{2} d\v q' \cdot (A(\v q') -A(\v q'+\v G)) \right\} + h.c.\nonumber \\
    &= -\frac{J}{2} \Psi_1\Psi^*_2\exp\left\{i\int_{1}^{2} +\int_{\bar{1}}^{\bar{2}} d\v q' \cdot A(\v q') \right\} + h.c.
    \label{HL12}
\end{align}
This resembles Eq.~\ref{SCH1}, but $J$ is a phenomenological parameter generated by integrating the degree of freedom on the 1D line. 

As an example, we explicitly derive the $H_{Fock}[\Psi_1, \Psi_2]$ for a generic potential $V(\v q)$. To do this, we start with Eq.~\ref{SCH1}, expand to $O(q^2)$ and carry out the $q$ integral,
\begin{align}
    H_{Fock}[\Psi]
    &= -\sum_{\v k}\sum_{\v q\ll K} \tilde{V}(\v q, \v k) \exp\left\{i\int_{\v k}^{\v k+\v q} d\v q' \cdot (A(\v q') -A(\v q'+\v G)) \right\} \Psi^*_{\v G}(\v k +\v q) \Psi_{\v G}(\v k)\nonumber \\
    &= -\int \frac{dk}{2\pi} \int \frac{dq}{2\pi} \frac{\tilde{V}(\v q, \v k) q^2 }{2} \left|\hat{\v n} \cdot \left(\nabla -i \v A(\v k)+i\v A(\v k+\v G)\right)\Psi\right|^2\nonumber \\
    &= - \int \frac{dk}{2\pi} \tilde{\tilde{V}}(\v k) \left|\left(\nabla_{\v n} -i A_{\v n}(\v k)+iA_{\v n}(\v k+\v G)\right)\Psi\right|^2
\end{align}
where $\hat{\v n}$ is the unit vector along the wire. We have denoted $\nabla_{\v n} = \hat{\v n} \cdot \nabla$, and $A_{\v n} = \hat{\v n} \cdot \v A$. We focus on the phase of $\Psi(\v k)=\rho(\v k) e^{i\theta(\v k)}$:
\begin{align}
    H_{Fock}[\theta]
    &= - \int \frac{dk}{2\pi} \tilde{\tilde{V}}(\v k) \rho(\v k)^2 \left(\nabla_{\v n} \theta - A_{\v n}(\v k)+A_{\v n}(\v k+\v G)\right)^2
    \label{SCH2}
\end{align}
To make progress, we ignore the $\v k$ dependence of $\tilde{\tilde{V}}$ and $\tilde{\rho}$. Then, integrate out $\theta(\v k)$ by demanding $\frac{\delta H}{\delta \theta}=0$,
\begin{align}
    \nabla_{\v n} \left(\nabla_{\v n} \theta - A_{\v n}(\v k)+A_{\v n}(\v k+\v G)\right)= 0
\end{align}
Under the boundary conditions $e^{i\theta(K_0)}=e^{i\theta_0}$, $e^{i\theta(\bar{K}_1)}=e^{i\theta_{\bar{1}}}$ (mind the $2\pi$ ambiguity in phase), this implies
\begin{align}
    \nabla_{\v n} \theta - A_{\v n}(\v k)+A_{\v n}(\v k+\v G) = \frac{1}{K}\left(\theta_{\bar{1}}-\theta_0+2m\pi-\int_{K_0}^{\bar{K}_1} d\v k \cdot \left( \v A(\v k)-\v A(\v k+\v G)\right) \right)
\end{align}
or
\begin{align}
    \theta(\v k)= \theta_{0}+ \int_{\v K_0}^{\v k} d\v k \cdot \left( \v A(\v k)-\v A(\v k+\v G)\right) + \frac{|\v k-\v K_0|}{K}\left(\theta_{\bar{1}}-\theta_0+2m\pi-\int_{K_0}^{\bar{K}_1} d\v k \cdot \left( \v A(\v k)-\v A(\v k+\v G)\right) \right) 
\end{align}
where $m\in Z$. Plugging in Eq.~\ref{SCH1},
\begin{align}
    H_{Fock}[\theta_0,\theta_{\bar{1}}]
    &= -\int \frac{d\v k}{2\pi} \int \frac{d\v q}{2\pi} \tilde{V}(\v q, \v k) \rho(\v k)^2 \exp\left\{-i \frac{q}{K} \left(\theta_{\bar{1}}-\theta_0+2m\pi-\int_{K_0}^{\bar{K}_1} d\v k' \cdot \left( \v A(\v k')-\v A(\v k'+\v G)\right) \right)\right\}\nonumber \\
    &= -\int \frac{d\v k}{2\pi} \int \frac{d\v q}{2\pi} \tilde{V}(\v q, \v k) \rho(\v k)^2 \exp\left\{-i \frac{q}{K} \left(\theta_{\bar{1}}-\theta_0+2m\pi-\left(\int_{K_0}^{\bar{K}_1} +\int_{\bar{K}_0}^{K_1}\right) d\v k' \cdot \v A(\v k') \right)\right\}\nonumber \\
    &= -\int \frac{d\v k}{2\pi} \int \frac{d\v q}{2\pi} \tilde{V}(\v q, \v k) \rho(\v k)^2 \exp\left\{-i \frac{q}{K} \Delta_{0\bar{1}}\right\}    
\end{align}
where we have defined $V_0(\v q)\equiv \int \frac{d\v k}{2\pi} \tilde{V}(\v q, \v k)\rho(\v k)$, and the gauge-invariant phase difference between two corners
\begin{equation}
    \Delta_{0\bar{1}} = \theta_{\bar{1}}-\theta_0+2m\pi-\left(\int_{K_0}^{\bar{K}_1} +\int_{\bar{K}_0}^{K_1}\right) d\v k' \cdot \v A(\v k')
\end{equation}
with $m$ taking the integer that minimizes $|\Delta_{0\bar{1}}|$. For concreteness, we explicitly evaluate the integral using a toy model with $V_0(\v q) = V\rho^2 e^{-g q^2}$, then
\begin{align}
    H_{Fock}[\theta_0,\theta_{\bar{1}}]
    &= -V\rho^2\int \frac{d\v q}{2\pi} \exp\left\{-gq^2-iq \frac{\Delta_{0\bar{1}}}{K}\right\}\nonumber \\
    &= -V\rho^2 \sqrt{\frac{\pi}{g}} \exp\left\{-\frac{\Delta_{0\bar{1}}^2}{4gK^2}\right\}\nonumber \\
    &\approx \frac{V\rho^2 \Delta_{0\bar{1}}^2}{4gK^2}\sqrt{\frac{\pi}{g}} + const.
\end{align}
We note that this leading quadratic term in $\Delta_{0\bar{1}}$ is expected for arbitrary $V_0(q)$, which can be shown by expanding Eq.~\ref{SCH1}.
%\begin{align}
%    H_{Fock}[\theta_1, \theta_2]
%    &= -\frac{V \rho^2 Q^3}{4\sqrt{\pi}K} \left(\theta_1-\theta_2+\int_{K_0}^{\bar{K}_1} d\v k \cdot \left( \v A(\v k)-\v A(\v k+\v G)\right) \right)^2\\
%    &= -\frac{V \rho^2 Q^3}{4\sqrt{\pi}K} \left(\theta_1-\theta_2+\int_{K_0}^{\bar{K}_1}+\int_{\bar{K}_0}^{K_1} d\v k \cdot \v A(\v k) \right)^2
%    \label{SCH3}
%\end{align}
%Comparing this with Eq.~\ref{HL12}, we identify $J=\frac{V Q^3}{\sqrt{\pi}K}$.
\begin{figure}[h]
    \centering
    \includegraphics[width=0.7\linewidth]{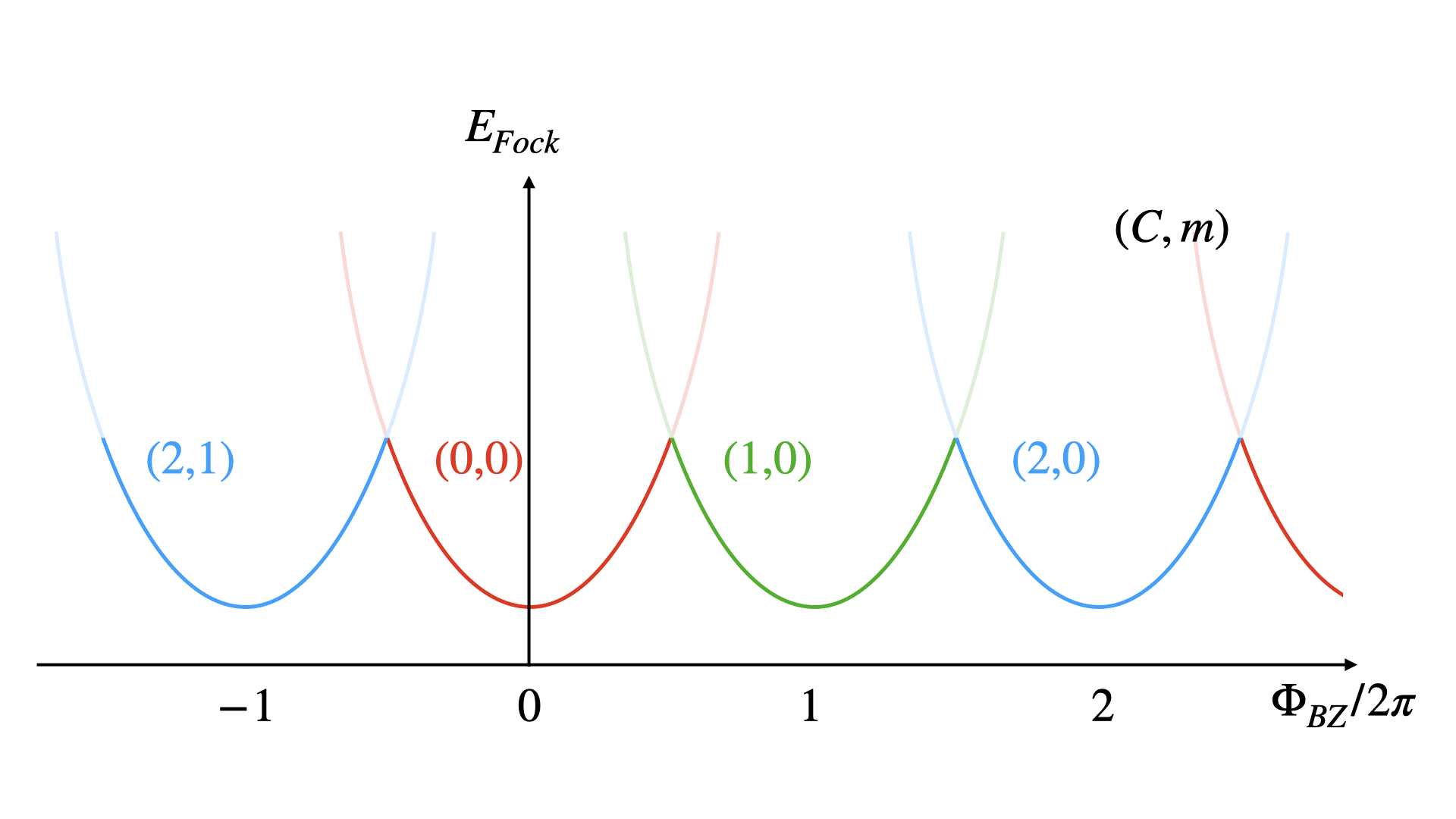}
    \caption{Schematic Fock phase diagram, the numbers label $(C \text{(mod 3)}, m)$ in Eq.~\ref{Eresult}.}
   \label{LittleParks}
\end{figure}
Using $C_3$ and combining the contribution from all edges of MBZ, similar to the discussion in Eq.~\ref{Chernnumber}, we identify the following 
\begin{align}
    C &= \frac{3}{2\pi}(\theta_0-\theta_{\bar{1}})\nonumber \\
    \phi_{BZ} &= -3\left(\int_{K_0}^{\bar{K}_1} +\int_{\bar{K}_0}^{K_1}\right) d\v k' \cdot \v A(\v k')
\end{align}
Therefore,
\begin{align}
    \Delta_{0\bar{1}} = -\frac{2\pi C -\Phi_{BZ}}{3} + 2m\pi
\end{align}
The $C_3$ symmetry demands the contribution from three edges to be exactly the same. Then, we find the Fock energy as a function of the Chern number
\begin{equation}
    H_{Fock} = H^{0\bar{1}}_{Fock} + H^{1\bar{2}}_{Fock} + H^{2\bar{0}}_{Fock} = 3H^{0\bar{1}}_{Fock}=\frac{V\rho^2\pi^{5/2}}{3g^{3/2}K^2} \left(C-3m-\frac{\Phi_{BZ}}{2\pi}\right)^2 + const.
    \label{Eresult}
\end{equation}
The phase diagram is plotted in Fig.~\ref{LittleParks}. 
We not only derive the rounding but also predict the Little-Parks effect: The Hall crystal/Wigner crystal becomes less stable when parent Berry flux approaches $(2n+1)\pi$, which may become useful when discussing physics at lower densities.
At reduced fillings and without a moir\'e potential, as we discussed in Sec.~\ref{sec_beyond}, the correlated fermi liquid is almost frozen, i.e. it behaves like a crystal at short distances and short times. It is natural to assume that  the WC/AHC short-range order has a lattice constant set by electron density. However, the picture in Fig.~\ref{LittleParks} suggests a route for the Fock energy to shrink the WC/AHC lattice so that the new BZ (with an area larger than the electron density) encloses a Berry flux quantized to an integer multiple of $2\pi$. The outcome of this competition between two crystalline orders is unclear to us, since the energy of the shrunk crystal also depends on the fate of the doped vacancies. We leave this to future study.

%At fractional fillings, there are two competing insulating states: (a) HC/WC with a periodicity consistent with the electron density (b) HC with a lattice matching with moir\'e, and fractionally doped. Both of them are commensurate with $V_{\text{moir\'e}}$ and therefore can be enabled even at intermediate interaction by a weak moir\'e potential. The Fock energy may favor (b) if the parent Berry flux in MBZ is close to $2\pi$.

\section{7-band Fock-only and Hartree-Fock phase diagrams}
\label{app:7band}
\begin{figure}[h]
{    \centering
\includegraphics[width=0.5\linewidth]{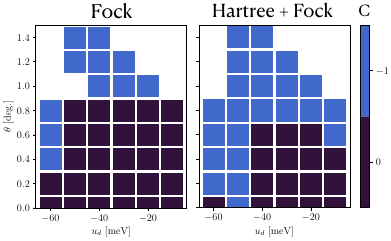}
 }       %\includegraphics[width=0.4\linewidth]{7_band_fock_only=18.pdf}
    \caption{Chern number of mean-field band obtained with Fock only [Left] and Hartree-Fock [right], respectively.
    These results are obtained from the 10-orbital model and the same parameters as Fig. \ref{hf_pd_compare}, with the exception of using 7-bands in the mean-field projection.}
   \label{fig_7_band_hf_comparison}
\end{figure}

We present in Fig. \ref{fig_7_band_hf_comparison} the Fock-only and Hartree-Fock phase diagrams computed with a 7-conduction band projection.
This provides a consistency (and convergence) check with our 4-band projection results presented in the main text.
As seen, the Hartree correction to the Fock-only phase diagram expands the region of $|C|=1$ in parameter space, in agreement with the results in the main text.

\section{Comparison of Fermi liquid and Hall crystal: Hartree Fock energies}
\label{app_fl_hc_comparison}
\begin{figure}[h]
    \centering
    \includegraphics[width=0.4\linewidth]{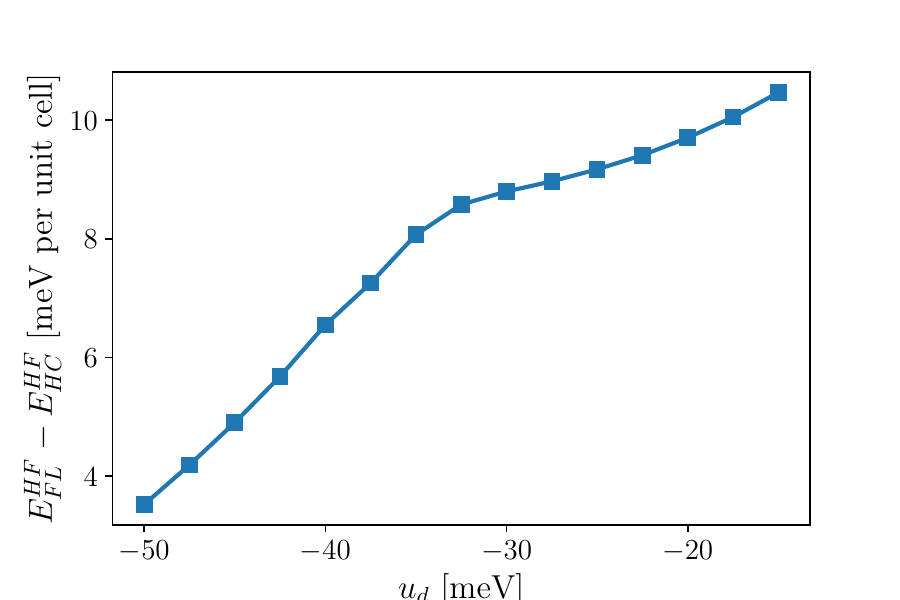}
    \caption{Energy difference of translation symmetry preserving Fermi liquid and translation symmetry breaking Hall crystal for wavevector $\vec{G}_M$ associated with \moire twist angle of 0.77$^{\circ}$ for $u_d = -36$ meV.}
   \label{fig_fl_hc_comp}
\end{figure}

We present in Fig.~\ref{fig_fl_hc_comp} the Hartree-Fock energy difference of the translation symmetry preserving Fermi liquid and translation symmetry breaking Hall crystal for wavevector $\v G_M$ associated with \moire twist angle of 0.77$^{\circ}$ for $u_d = -36$ meV.
As seen, for a range of displacement field energies, the Fermi liquid state is higher in energy than the Hall crystal.
As alluded to in the main text, this energy difference is not altogether surprising, as the Hartree-Fock framework is biased towards finding translation symmetry broken (and gapped) phases of matter.

\end{document}